\documentclass[preprint, amsmath,amssymb,aps,prper]{revtex4-2}

\usepackage{graphicx}
\usepackage{dcolumn}
\usepackage{subfigure}
\usepackage{amsthm}
\usepackage{bm}
\usepackage{multirow}
\usepackage{longtable}
\begin{document}
\title{Importance of accounting for student identities and intersectionality for creating equitable and inclusive physics learning environments}
\author{Lisabeth Marie Santana}
\author{Chandralekha Singh}
\affiliation{Department of Physics and Astronomy, University of Pittsburgh, Pittsburgh, PA, USA 15260}

\date{\today}

\begin{abstract}
    This research focuses on the experiences of seven undergraduate women who were majoring in physics in a medium-size physics department at a small liberal arts college. 
    In the semi-structured, empathetic interviews we conducted, the women discussed how they decided to major in physics, their interactions with their peers and instructors, who supported them during their physics trajectory, and suggestions that would improve their experiences in physics.
    We used Standpoint theory and focused on the experiences of undergraduate women to get a holistic perspective of how they became interested in physics, how they have been supported in their physics journey as well as identify any challenges that they faced in their undergraduate physics program due to their identity. 
    Using synergistic frameworks such as the Domains of Power and the Holistic Ecosystem for Learning Physics in an Inclusive and Equitable Environment (HELPIEE), we analyzed how those in the position of power, e.g., instructors, can play important roles in establishing and maintaining safe, equitable, and inclusive environments for students, which is especially important for historically marginalized students such as women and ethnic and racial minority students in physics. 
    We also discuss the suggestions provided by the undergraduate women to implement in the future to support current and future undergraduate women in physics and astronomy. Their suggestions are separated as personal advice for peers and suggestions for physics instructors.

\end{abstract}

\maketitle

\section{Introduction}

Many groups, e.g., women and ethnic and racial minorities (ERM), have historically been underrepresented from physics and improving their representation and experiences has been an important focus for researchers for the past few decades. 
Within the last decade, the percentage of women who received physics bachelor's degrees in the US has 
ranged from 20\% to 25\% \cite{AIPbachelorsdegrees2018, AIPbachelorsdegrees2020, APS2020PhysicsDegrees}. 
Reports from 2020 shows that of physics bachelor's degrees, 3-4\% are awarded to Black students and 12\%  are awarded to Latinx students \cite{AIP2020Physicsdegrees, boatman2022}.

Moreover, the women and ERM students who obtain their degrees in physics, face many challenges while doing so uniquely due to their identities. These issues faced by women and ERM students in many undergraduate Science, Technology, Engineering and Mathematics (STEM) fields have been reported in previous studies \cite{seymour1997talking, seymour2019talking, pollack2016, hazari2007genderdifferences, sax2016wip, potvin2015underpresentation, hazari2020context, lorenzo20066gendergap, Cheryan2017STEMGender, jones2000gender, hill2010few, ganley2018gender, mccullough2011women}. 
Research focusing specifically on physics and astronomy shows that women and ERM students 
often have negative experiences in the physics learning environments \cite{barthelemy2016gender, traxler2016genderinphys, rosa2016obstacles, santana2022investigating,santana2023effects} and there are often gaps in performance and psychological factors (e.g., self-efficacy, belonging, identity etc.) disadvantaging traditionally marginalized students.

At many institutions, physics is still tied to its historical origins in regard to its culture. For example, physics has stereotypes among the natural sciences regarding  who belongs in it, who can excel in it and what a physicist looks like \cite{leslie2015expectations, santos2017you}. 
Students are negatively affected as a result of these stereotypes \cite{rosa2016obstacles}, including in their performance 
\cite{marchand2013stereotypes, cwik2021damage, maries2018agreeing}. In addition, the demographics of physics departments may continue to reinforce stereotypes, meaning historically marginalized students, who don't have role models, continue to be constantly reminded that they do not necessarily fit in.

Previous studies have shown gender disparities in performance as well as psychological factors in physics persisting throughout high school and college. Such disparities include women's lower perceived recognition compared to men as a ``physics person" from peers and instructors \cite{hazari2010connecting, kalender2020damage, kalender2019female, nissen2016gender, zeldin2008comparative, lock2015physidentity, cwik2022not, liimpact2023}, which can hinder their academic performance \cite{cwik2021perception} and influence women's decisions to leave the field \cite{sawtelle2012exploring, seymour1997talking, seymour2019talking, good2012optout}.
In particular, if women believe that their instructors or peers do not see them as being capable of excelling in physics, it can impact their own beliefs about whether they can excel in physics \cite{li2020perception,li2021effect, lock2015physidentity, hazari2015powestructures}. Moreover, lower self-efficacy and sense of belonging can be detrimental even to those students who are performing well. Prior research also shows that women drop out of STEM disciplines with significantly higher grades than men \cite{maries2022gender}. Prior studies have also found that sense of belonging and self-efficacy in physics are closely intertwined \cite{kalender2019gendered, lewis2016fittingin, doucette2020hermione}. Because improving self-efficacy is a multifaceted process \cite{bandura1999self}, creating an equitable and inclusive learning environment that increases underrepresented students' sense of belonging may help increase their self-efficacy and improve their retention \cite{masika2016building, goodenow1993classroom, marshman2018female, marshman2018longitudinal, raelin2014gendered, atwood2010ac,whitcomb2020comparison, felder1995longitudinal}.

We also note that physics culture has traditionally been masculine, as it still is at many universities, which continues to perpetuate inequities and stereotypes for traditionally marginalized students. The physics culture proclaims and reinforces ideas that only brilliant people can do physics and since men are stereotypically associated with brilliance \cite{bian2018brilliance}, they are the only ones who can do physics and belong in physics \cite{pollack2016, allen2016women, maries2018agreeing, maries2020active, karim2018evidence, walton2015two, steele2010stereotypes, reuben2014stereotypes,  marchand2013stereotypes, kelly2016gender, mccullough2007gender, Gonsalves2016Masculinities}. This masculine culture disadvantages women and they become isolated from lack of role models and a community that could provide support to them \cite{francis2017construction, danielsson2012exploring, Gonsalves2016Masculinities, dennehy2017mentors}.

It is important to investigate the experiences of underrepresented students in physics, such as women as in our study, since they are the ones who can shed light on their interactions with others in physics, and how the physics culture impacts them.
Qualitative studies such as this investigation can be valuable to learn about inequities directly from students who may be experiencing it.
Previous studies involving large research universities in the US in other contexts that utilize student interviews revealed that women in physics and astronomy experience hostile environments within these physics departments \cite{barthelemy2016gender, doucette2020there, rosa2016obstacles, walton2015two, santana2022investigating, santana2023effects}. 

In the US, there are many small liberal arts colleges with physics departments of various sizes (with regard to the number of faculty members and physics majors), which may potentially have very different physics cultures even if they are all liberal arts colleges. Moreover, the physics culture and its impact on traditionally underrepresented students such as women at these small institutions with physics departments of different sizes can be different from that at the large research universities that have been investigated in most prior qualitative research studies from the introductory to graduate levels. We also note that apart from Johnson's study \cite{johnson2020intersectional} in a small physics department at a small liberal arts college,
we are not aware of prior qualitative studies that focus on the experiences of women physics majors in physics learning environments at small liberal arts colleges with physics departments of any size in the US. Also, to our knowledge, Santana and Singh qualitative study is the only one focusing exclusively on the experiences of undergraduate women majoring in physics \cite{santana2023effects} at a large research university in the US.  

Furthermore, Kanim and Cid have emphasized that physics education researchers should design and implement studies at a variety of institutions with physics departments of different sizes so that the research is representative to the broader physics community and benefits students in different types of institutions and physics departments \cite{kanim2020demographics}.
This investigation is an important step towards including the experiences and perspective of women physics majors in a 
medium-size physics department at a small liberal arts college. This study also provides opportunity to compare women physics majors' experiences in a medium-size physics department with a small physics department in Johnson's study \cite{johnson2020intersectional} and a large research university \cite{santana2023effects} in the US in prior qualitative studies to better understand the similarities and differences in women physics majors' experiences 
in physics departments of different sizes at various institutions.
Through interviews, we can understand undergraduate women's perspectives and their experiences in physics learning environments in their own voices.

This research uses qualitative data to investigate the experiences of undergraduate women majoring in physics in a medium-size physics department at a small liberal arts college in the US.
In particular, we use qualitative interview data to understand undergraduate women's trajectory in physics (including what or who got them interested in physics), interactions with their peers and instructors, who supported them in physics and challenges they face, and their suggestions to improve their experiences in the physics learning environments in a medium-size physics department at a small liberal arts college in the US.

Also, although a detailed comparison is out of the scope of this paper, we summarize how the findings at the medium-size physics department (presented in detail here) compare to Johnson's study in a small physics department at a small liberal arts college and a study we conducted 
at a large research university to analyze if there were major differences in women's interactions with male peers and faculty across these different types of institutions \cite{johnson2020intersectional, santana2023effects}.

\section{Frameworks}
\subsection{Standpoint Theory}
We choose Standpoint theory as our theoretical framework to help frame this study in which we put undergraduate women majoring in physics at the forefront \cite{rolin2009standpoint,longino1993feminist}. Standpoint theory is a critical theory that focuses on the relationship between the production of knowledge and acts of power \cite{harding2004feminist}. It is related to other critical theories in that it centers around the standpoint or voices of the  underrepresented groups that do not have the same privilege as the dominant group in order to gain a clearer understanding of their struggles.
Thus, using this framework, we put emphasis on the experiences of undergraduate women in physics to understand what physics departments and instructors can do to improve the physics culture such that they feel safe and supported \cite{blickenstaff2005leakypipeline}. It is also important to acknowledge that women are experts in their own experiences, thus these interviews can provide valuable insight about how they navigate through various situations with this specific lens. From these interviews, we also obtain a lot of valuable information to improve the current physics learning environment, such as if and how they feel supported or what things must be improved upon.
For example, these types of experiences can be used to inform/educate those in the position of power about how the women perceive dominant groups, or to reflect upon and evaluate one's department to restructure the physics program/support systems.
Thus, using this framework, we chose to interview only undergraduate women in physics in this medium-size physics department at a small liberal arts college.

\subsection{Domains of Power}
Collins introduced four interconnected domains that are necessary to understand how power is organized in a particular context \cite{collins2009another}. She argued that the four domains we need to consider are: interpersonal, cultural, structural, and disciplinary. The interpersonal domain pertains to how power is expressed between individuals. The cultural domain pertains to how group values are expressed, maintained or challenged. The structural domain pertains to how power is organized in various structures. The disciplinary domain pertains to how rules are enforced and for whom. Johnson applied this framework in the context of physics learning environments in a small physics department at a small college where women of color felt supported and were excited about learning physics. Within the Domains of Power framework, it is those in the position of power, e.g., instructors, who have the power over shaping the narrative for each of these domains (which interact with each other) in the physics learning environments \cite{johnson2020intersectional}.
In the context of student experiences in learning physics, the interpersonal domain refers to how students communicate with each other and how students and faculty interact; the cultural domain refers to the physics culture;  the structural domain refers to the structure of a physics learning environment; and the disciplinary domain refers, e.g., to how instructors discipline students in the physics courses and in the physics learning environments in general, if their conduct does not conform to the expected norms.

\subsection{HELPIEE}
\begin{figure}[tb]
\small
\begin{center}
\includegraphics[width=0.6\textwidth]{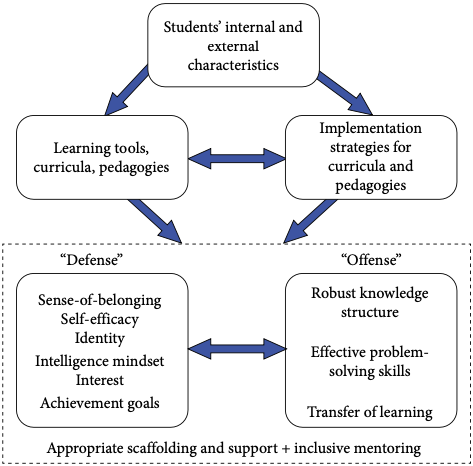}
\end{center}
\caption{Figure from \cite{cwik2023framework}, summarizing the HELPIEE framework.}\label{fig:helpiee}
\label{helpiee.fig}
\end{figure}

The HELPIEE framework posits that those in the position of power, e.g., instructors in physics classrooms, have the power to help all students feel supported by carefully taking into account students' characteristics and implementing effective pedagogies in an equitable and inclusive learning environment \cite{cwik2023framework}.
Student learning of physics can be affected by students' backgrounds including their internal and external characteristics and whether learning environments are equitable and inclusive and provide appropriate support to all students based upon their needs. Internal characteristics include prior preparation, physics self-efficacy, identity etc. and external characteristics include their other academic and non-academic commitments and support systems such as the availability of appropriate advising/mentoring etc.
Figure \ref{fig:helpiee} summarizes the HELPIEE framework in which accounting for students' characteristics can help instructors intentionally focus on student needs and tailor how to adapt and implement pedagogies 
in their courses to strengthen both students' ``Offense" or ``Defense" skills for learning physics.
Within this framework, if students from different demographic groups in a course, e.g., those from underrepresented groups, do not have similar positive experiences and feelings of being supported, the learning environments are not equitable because those in the position of power did not provide adequate support to level the playing field.

Both the Domains of Power and HELPIEE frameworks are synergistic. For example, in the context of physics courses, it is the department and instructor's responsibility to create an equitable and inclusive learning environment.
Together, these synergistic frameworks are useful for analyzing the findings of this study and comparing them with prior studies. Standpoint theory aligns with the two by putting the emphasis on women's experiences to reveal the types of interactions they have in physics learning environments, what the physics culture is like, and what instructors' actions are (or aren't) in creating equitable learning environments. Thus, Standpoint theory acts as an overarching lens to focus on and learn from undergraduate women majoring in physics, who share their experiences.
We also note that Standpoint theory also helps in the empathetic interviews by allowing the interviewer to connect with the participants based on this shared identity.

\section{Methodology}
\subsection{Participants}
We conducted semi-structured, empathetic interviews with 7 undergraduate women physics and astronomy majors in a medium-size physics department at a small liberal arts college in the US. At this college, women are underrepresented in physics, but not to the extent that they are underrepresented at the national level \cite{AIPbachelorsdegrees2018}, i.e., they make up about one-third of the physics majors.
We note that there are 9 full-time faculty members and 4 visiting/adjunct faculty members in this physics and astronomy department. The number of physics majors that graduate each year varies from 10-20.
In order to capture a wide range of experiences, we interviewed women who were at various points in their physics trajectory in college, e.g., second to fourth year students. 

CS contacted a faculty member at this college who then sent an email to all 20 undergraduate women (across different years) who were majoring in physics at the time. The email described the scope of the study and asked for those interested in participating to reach out to CS. Each of the seven participants (approximately one third of the women in that undergraduate physics program) volunteered to participate without compensation. 
At the beginning of each interview, the participants provided consent to be audio-recorded and quoted in academic publications. 
Each interview was about an hour long in duration and was recorded via Zoom. The interviews followed protocols set by the group prior to conducting the interviews (see Table \ref{tab:questions}). The protocol questions focused on the women's experiences related to physics, such as what their high school experiences were in STEM, what their interactions were like with college peers and instructors, and whether and how their gender may have impacted their experiences. The interviews were semi-structured and we followed up on what the women described in response to the questions that were already part of the protocol.

We briefly introduce each woman in this study who was majoring in the physics and astronomy department in a medium-size physics department at this small liberal arts college in Table \ref{tab:intro_women}. All women were provided pseudonyms for anonymity. Several of the women (Cathy, Francis, and Karina) were either double majoring or interested in other subjects, such as computer science, mathematics, or English. All of the women were brought up in the US, except Francis who grew up in Europe and came to the US to study at this college. 

\begin{table}[tb]
\caption{\label{tab:intro_women}%
Here we provide information about the women that were interviewed.}
\begin{ruledtabular}
\begin{tabular}{lll}
Pseudonym & Ethnic / Racial Background &  Year in college \\
\hline
Cathy & White & 4th year \\
Chloe & White & 3rd year \\
Esme & Latina & 2nd year \\
Francis & White & 3rd year \\
Karina & White & 4th year \\
Paulina\footnote{We note that Esme and Paulina have different perspectives regarding which of their marginalized identities was prominent in their negative experiences. One reason for this could be the differences in their visual characteristics, i.e., only Paulina visually appears to be non-White.} & Latina & 3rd year \\
Ruby & White & 2nd year \\
\end{tabular}
\end{ruledtabular}
\end{table}

\subsection{Positionality}
Both researchers identify as women of color. Specifically, LMS identifies as a queer Chicana woman and CS identifies as an Asian American woman. It is important to note these identities as they are also intersectional and were useful when coding and analyzing the intersectional identities of some of these undergraduate women in physics. Both have had several prior experiences conducting and analyzing qualitative studies focusing on equity and inclusion in physics learning environments.

\subsection{Data Analysis}
We coded these interviews using hybrid coding (inductive and deductive) methods that emphasized the women's experiences \cite{saldana2013coding}. Deductive codes are based on the interview questions while inductive codes were formed during the analysis \cite{saldana2013coding}. For example, based on the interview questions (see Table \ref{tab:questions}), we anticipated that the women would provide suggestions to improve their experiences although we did not anticipate the women to describe such specific suggestions that could be separated as personal advice for other women and advice for instructors. Another example of an inductive code was the reflections women provided regarding how their identities other than gender influenced how they navigated physics learning environments. 
Again, these themes emerged and were not directly based on the protocol questions specifically focusing on these.

To begin the analysis, LMS first corrected all the interviews from the Zoom transcriptions by listening to the audio-recordings to ensure that no information was misinterpreted. Both researchers read all the transcripts and had many extensive discussions about them with each other. Then, after these discussions between the two researchers, LMS proceeded to code the interviews using fine grain, specific codes, starting with one woman at a time. During this 
round, LMS and CS again engaged in several discussions about the codes including definitions and clear examples from the data. 
Through many back and forth discussions and fine-tuning, they came to an agreement on the general codes extracted from the fine grain codes. The general codes were then distilled into subthemes, and corresponding  analytic themes \cite{saldana2013coding}. In general, several subthemes were part of the same analytic theme. The researchers discussed and justified these subthemes and analytic themes based on how broad they were 
in elucidating the women's experiences (capturing both negative and positive interactions with peers and instructors).

\begin{figure}[tb]
\small
\begin{center}
\includegraphics[width=0.5\textwidth]{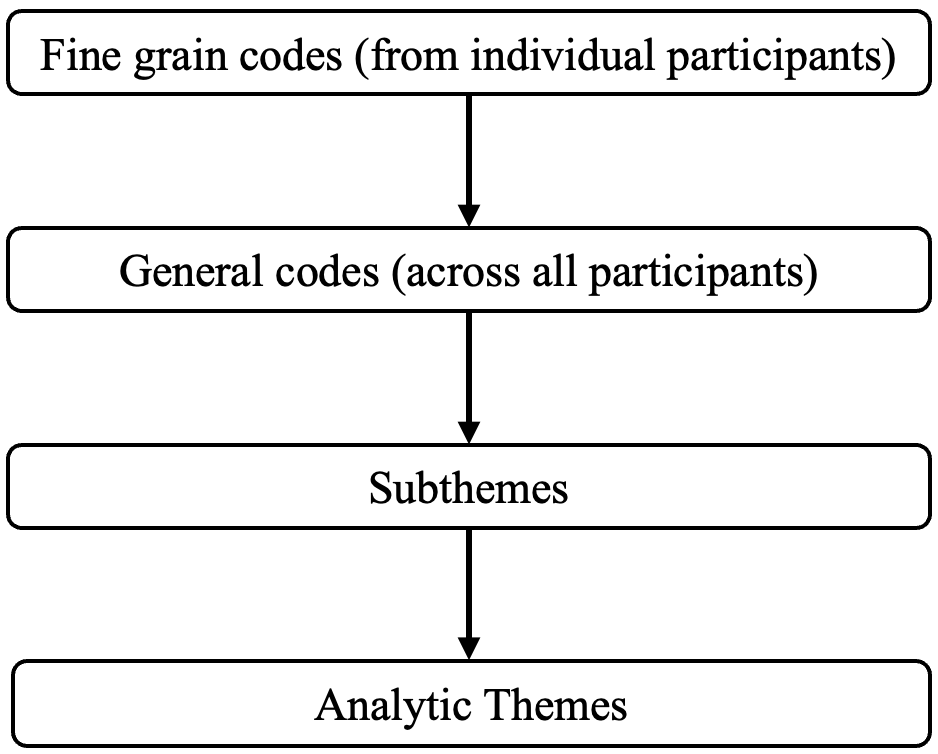}
\end{center}
\caption{Layout of the data analysis, adapted from Figure 1.1. in \cite{saldana2013coding}}\label{fig:theme}
\label{theme.fig}
\end{figure}

\begin{figure}[tb]
\subfigure[]{\label{fig:GC1} \includegraphics[width=0.7\linewidth, height=9cm]{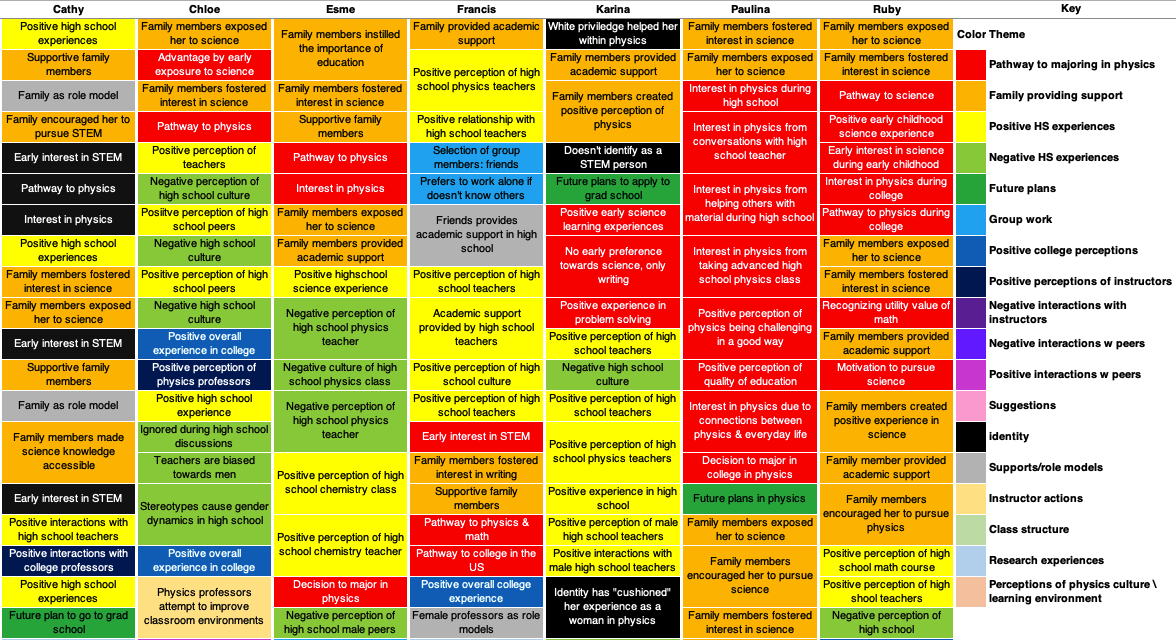}}
\subfigure[]{\label{fig:GC2} \includegraphics[width=0.7\linewidth, height=9cm]{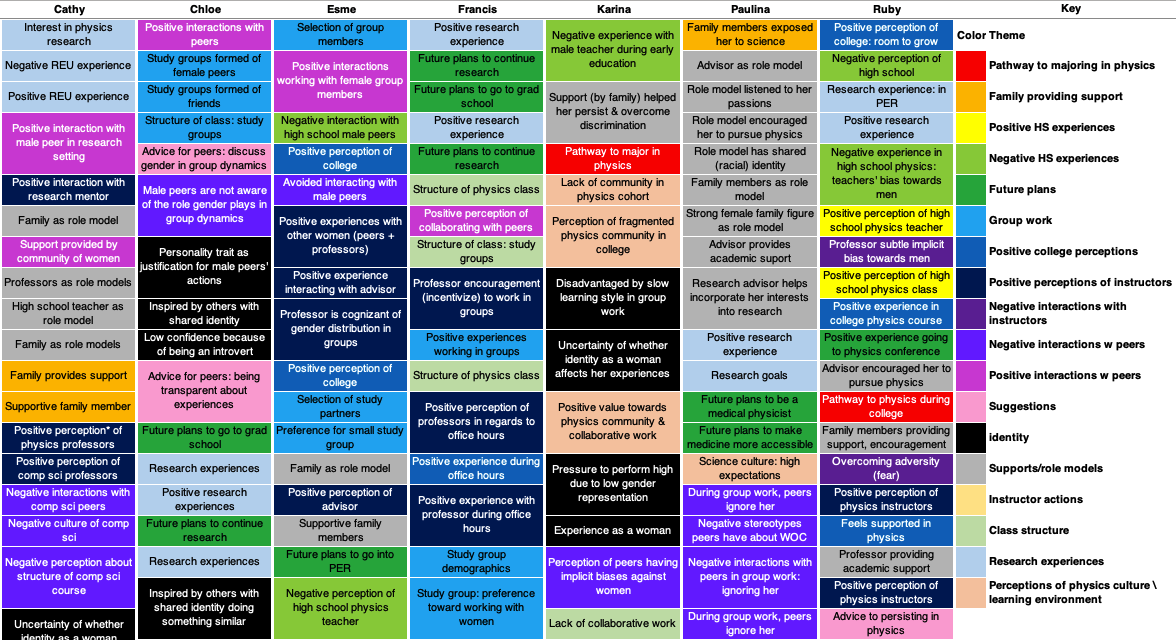}}
\caption{Examples of general codes from participant data. Each color refers to a separate theme. The key lists abbreviations in some cases. Some of these themes were not discussed in this paper.}
\label{fig:GC}
\end{figure}

Figure \ref{fig:theme} provides a layout for how the codes were distilled starting from fine grain codes from individual participant data, to analytic themes.
Fine grain codes are very specific for each participant and are not included here.
In particular, depending upon their experiences, the women might have had similar or different fine grain codes for a particular question, which were then grouped 
into general codes. 
Thus, general codes
are broader than fine grained codes to include data from all of the women.
Table \ref{tab:examples} provides general codes, their definitions, and examples from individual participant data.
Next, the general codes were distilled into subthemes. 
As shown in Table \ref{tab:examples}, these subthemes are broader than the general codes.
Also, for illustration purposes, Figure \ref{fig:GC} shows a subset of general codes which we color coded
for a corresponding subtheme agreed upon by both researchers.
As shown in Table \ref{tab:examples}, we grouped similar subthemes under Analytic Themes. 
We note that only one Analytic Theme (Reflection on How Identities Affect Experiences) did not have subthemes so that the subtheme and analytic theme are the same in this case.

As an example, Table \ref{tab:examples} shows that for the first Analytic Theme ``Pathways to Majoring in Physics", one subtheme was
``Family's role in providing support''. The general codes for this were: Providing academic support, Providing encouragement to pursue science, and Providing exposure to science. Depending upon the narratives of the women, there were slight variations in their fine grain codes under this general code.
We note that in the excerpts, we removed most of the ``ums" and ``likes" to make it easier to follow their point of view.
We also note that occasionally some women elaborated more than others regarding certain instances, so some voices may come across as more emphatic pertaining to a particular issue than others.

\subsection{Analytic Themes}
The main analytic themes (AT) related to their experiences that we focus on are the following:

\begin{description}
    \item [AT1]Pathway to Majoring in Physics
    \item [At2]Interactions with College Peers
    \item [AT3]Interactions with College Instructors
    \item [AT4]Sources of Support
    \item [AT5]Reflection on How Identities Affect Experiences
    \item [AT6]Suggestions to Improve Experiences
\end{description}

\section{Findings and Discussion}
Below, we present the findings from undergraduate women in physics at this institutions. The analytic themes are presented following their subthemes and general codes \footnote{General codes are not always bolded in the following sections because they were intertwined in the data.} (bolded).

\subsection{Pathway to Majoring in Physics}
In this section, we discuss the experiences that influenced the women's decision to major in physics in college, e.g., family's role, and positive high school experiences, while also discussing their negative high school experiences.

\subsubsection{Family's role in providing support}
All of the women described several ways in which their family provided support to them, e.g., providing general support, academic support, and exposure to science at an early age. These positive experiences with family as described below were very important in putting the women on the path to majoring in physics and potentially developing a science identity.

\textbf{Family providing academic support}: Many women described how family members supported them in their physics journey through having first hand experiences with related disciplines themselves. For instance, some described that family members were physicists, majored in physics, or had STEM degrees, which required taking physics courses.

Cathy describes how her sister, who is doing her Ph.D. in physics, helped her learn physics and provided encouragement to major in physics:
``I kind of like learned a lot about it from her and she encouraged me to go into it...I think that she made it a lot easier for me to go into it...she had a lot of good advice and...just kind of made the whole thing like pretty accessible.''
Similarly, Francis explains how one of her older sisters who majored in engineering has provided academic support. For example, she is able to go to her sister for homework help.
Here we see that having someone directly in the field was very beneficial to Cathy and Francis.
In addition, receiving advice from an insider, someone who already went through the process, may have been especially beneficial as that advice may have been specific to physics because even within different STEM disciplines, the established norms and cultures can be different.

Chloe also describes how her parents being physicists have influenced her decision to major in physics. She says,
``I've always been surrounded by [physics], so I think I almost had an advantage in that way, in which seeing actual physicists and particularly female physicists-it had a big influence on me and I've always kind of wanted to pursue physics, particularly astronomy...''
Being surrounded by physicists was beneficial to her, e.g., not only is her self-efficacy likely to get a boost particularly because her mother is a physicist but close family ties with physics possibly introduced her to various topics in physics and astronomy early on and provided insights as to what a career in physics may look like. In particular, seeing female physicists may have been especially beneficial in that Chloe was seeing people with a shared gender identity pursuing and being successful in a field that she had interest in.

Karina also explains how her parents provided academic support because they are electrical engineers:
``...I enjoyed physics in high school, but I also had parents who understood it, so if I was confused about something, you know, it was similar to a lot of work they were doing as electrical engineers, so I could go home [and] ask my parents, \textit{Oh I don't understand how this works}, and I think kind of like having that support and getting the early on understanding definitely helped me enjoy physics and feel like I yeah could get it.''
We see that having family members who are familiar with physics as they had taken physics courses, can be especially helpful and encouraging to women like Karina.

\textbf{Family providing exposure to science}: A few women, such as Esme, Paulina, and Ruby, have family members who exposed them to science broadly either through their career or through experiences not related to school.

Esme describes how her dad instilled the importance of education into her:
``He works in construction and...since he was born in Mexico, he couldn't get an education here in the US, so he kind of always projected [that] education is important. It was the number one thing I had to focus on and he was always into engineering and all that stuff and because we were so close, that's how I got into physics. I've always been into astronomy. So he had always...encouraged me to go into whatever I liked and I've always been very STEM oriented because of him.''
Thus, Esme's dad encouraged her academic endeavours and his interest in engineering influenced her own interest in physics.

Esme also explains that when she was young, her dad would take her to work and expose her to the technical aspects of his job:
``he kind of would always explain how cars work or because he would work in construction...sometimes, he would take me when he wouldn't do the dangerous things and he would just explain what he was doing...I remember this one thing-we went to the library and we just got a book on the history of math and we just looked at the photos because I was very young and he would just explain to me all the...optical illusions and stuff like that...''
These are key experiences for Esme as her dad was both encouraging and introducing her to STEM related topics, e.g., those that were related to his job or exposing her to other science and math topics, e.g., optical illusions. We also see the effort by her dad to engage Esme in STEM experiences in a supportive manner.

Paulina shares that her family having careers in STEM exposed her to science:
``I was always interested in STEM because a lot of people in my life were either doctors and so they did STEM for a living...[or] my father and my grandpa, for example, they were both chemical engineers, so I'd always been like surrounded by STEM...most of it had been from a very young age.''
This early exposure may have fostered Paulina's interest in science. She directly credits her interest in STEM to having family members in STEM. In addition, she feels very good about the fact that she was ``surrounded by STEM" so these various exposures appear to have instilled interest in STEM disciplines.

Paulina also shares how her grandmother taught her different aspects of the human body and medicine and interacting with doctors provided additional exposure to medicine:
``whenever I was spending time with [my grandmother]...she would either give me small little pointers or small literally tidbits of medical facts...and be like, \textit{Hey, what is the femur? What is the function of it in the body?} and simple stuff like that to the point where I became more engaged with medicine, at least, and I would also use that a lot to interact with doctors...and so I would use information my grandmother would tell me, her little facts and then eventually I was able to communicate with doctors a lot and whatever kind of information they told us about a diagnosis. I would also retain that because, again, I was one of the only people [in] my family that was able to speak English and Spanish, and so I was often in charge of communicating to my family members and doctors. So whatever information I got to hear about medicine in Spanish, I would translate that and then I would also get some other information from doctors. And so that happened very frequently in my life, and that was how I became really, really immersed in STEM.''
Her interview suggests that this constant exposure may have provided a sense of identity in science as Paulina was able to serve as a translator between her family and doctors.
She further explains that this exposure to scientific information was repeated throughout her childhood like a cycle. It appears from her interview, that this continuous exposure to science may have been useful in creating her science identity.

Paulina also recalls times with her grandfather in which he would explain the machinery he used and taught her about chemistry:
``he kind of showed me around the site where he was working and he told me the basic functions of a lot of the machinery that he was using and most of his work involved chemistry, so I was often listening to chemical composition of a lot of materials so it wasn't necessarily physics, but I was always engaged with all these different aspects of STEM all around me. And I think he kind of just stuck with me as I grew up and I just realized that I really wanted to keep learning about it because it's something that [I] had been so familiar with for a long time.''
These different exposures to science made her want to continue learning and possibly continue pursuing science beyond high school.

Ruby explains that although many of her family members are not scientists, her grandfather provided exposure to science:
``So actually my parents are both not scientists and I have three brothers who are also not in science as well, but my grandpa was very into science and at a young age, he would always do these math puzzles with me and I just found [it] really interesting because it was like, you have a step by step approach of how to get to a solution.''
In this early recollection, she describes how doing math puzzles introduced her to problem solving. Additionally, she elaborates: ``but he would give me numbers and then there was an equation that I had to get to using the numbers that he would give me and he would have some hints for how to get there, but then the rest I would have to figure out on my own and then use those steps that he gave and add on to them to get to the final product.''
She summarizes that having her uncle as a physicist and her grandpa who was into science was helpful: 
``So I think having those two people in my life, who were very active in sciences and in a way, having my parents, not being in science kind of made me realize I want it for me, not for any other reason.''
Thus, for Ruby, having family members who were both into science and not in science seemed beneficial as she had several perspectives and several family members who helped her get into physics as a college major.

We see the role family has had in supporting the women majoring in physics at this liberal arts college. Not only did family provide academic support, if they had careers in STEM, but also provided opportunities for the women to interact and learn science in contexts that were applicable to their jobs or pursued scientific activities for fun, e.g., Ruby's puzzles with her grandfather.

\subsubsection{Positive high school experiences}
Many of the women took a physics class during high school. We recount their positive experiences in their high school physics class. These positive experiences capture a range of interactions and perceptions of their peers and/or teachers. For example, they noted that the teachers were supportive and peers were driven, which inspired them.

Cathy explains that the high school she attended was a `pretty liberal school' and that she liked her high school teachers:
``I was pretty lucky because everything was pretty gender balanced and they really encouraged women to pursue STEM. It was definitely one of their focuses.''
It seems like Cathy's teachers were generally supportive and encouraging. When asked if she experienced any gender bias, she says,
``I didn't feel that in high school, but I've definitely felt that in college.''
In comparison to her college experiences, which we will discuss in a later section, her high school experiences were generally positive.

Esme attended a public school and shares that she was part of a NASA program during her junior year.
``I got to go to NASA for a week, all expense paid trip and I would be there and...we did like a mock Mars mission and I got to work with actual engineers and they would give us lectures and give us tours around NASA, it was very great opportunity...''
This positive experience may have solidified her interest in physics and astronomy by allowing her to work with NASA engineers and learning about astronomy.

Francis attended an all girls high school in a European country. She didn't feel out of place because, 
``it was just all girls...''
She also describes having ``nice physics teachers.''
For example, she explains that when it came to college applications, her physics teachers were, 
``always very willing to help with applications and interviews.'' She explains that in her country of origin, when applying for a college degree, some universities will have interviews. Francis also describes how her physics teachers would help by doing mock interviews in which
``they would ask physics questions and solve it and talk through how I solved it and they would definitely [be] very helpful with that.''
Thus, Francis' physics teachers played a very important role in helping her prepare for college, in addition to encouraging her and other students to apply for college.

Paulina recalls her interest in physics and taking several classes during high school. Through this class she was able to have a lot of hands-on experience in labs and was able to have one-on-one conversations with her teacher:
``I became really interested in the class and I would stay after [to] chat with him, ask questions, help other students and I think at that point, I just kind of viewed it as a hobby of mine. I didn't really expand on it too much and so after that I proceeded to take AP physics C... electromagnetism and mechanics and that was also very interesting to me. I think it was one of the first experiences that I had where I felt truly challenged in a class, but also in a very positive way. I felt I was getting something out of the education I was receiving and it was very interesting, especially because I could visually see the ways in which physics and I interacted a lot in my everyday life. And that was what really first got me into physics.''
It is possible that having these one-on-one conversations with her high school teacher was beneficial and helped her see physics as an interesting field to pursue, and potentially a field in which she can have a career instead of pursuing it only as a hobby. These conversations may have also encouraged her to continue challenging herself, such as by taking the AP Physics exam. Her perception of physics throughout the interview was very positive and she found connections between physics she learned in class and physics she interacted with in everyday life.

Chloe attended a private school which she says was known for being a
``competitive college prep school where there are many students who had been like training in math Olympiads and science competitions their whole lives.''
She had overall mixed experiences during high school, but in this section we will comment on her positive experiences. Despite the negative competitive culture, she says, 
``...it was inspiring to be around people who were so driven academically, so I think I grew academically... I wouldn't have if I hadn't been surrounded by these people who are just really focused on getting into college.''
The fact that students at her high school were driven and focused on college education encouraged her to also focus on her academic trajectory.

In regard to her experiences in her high school physics classes, Chloe said that in her AP physics class, women made up half of the class and that, 
``...all these girls were like top of our class, really wanted to go into STEM, so I had a positive experience with that because they're also the people I was friends with. So it was nice being in a class where I had friends there and then they're all very driven.''
Being surrounded by other women who also wanted to pursue STEM created a positive culture/environment for Chloe. It also helped that she was also friends with these peers.

Karina also attended a private high school and had mixed experiences. Her positive experiences were mainly in regard to her high school teachers. When asked about potential gendered experiences in high school, she says,
``Both teachers were male, but they were very supportive, and I don't think I ever felt treated unfairly because of my gender...I always felt they treated me like I was smart and capable and deserved to be there...I can definitely remember for that advanced physics class my senior year, there was a big gender disbalance, and I think even if people are treating you fairly, if you feel like one in the minority, it does affect your experience."
Despite there being more men than women in her high school physics class in her senior year, Karina still felt supported by her teachers. She also adds an important fact that having an underrepresented identity can affect a student's experience even if others are treating them ``fairly".

Ruby also had mixed high school experiences. She says she liked her high school teachers and notes that she had positive experiences in her math classes. Specifically, she describes her high school physics teacher:
``In terms of asking him for help, or questions, I did find that he was very approachable and he wasn't condescending or more condescending towards a girl [than] he would be towards a guy, which was really nice, and I think I never felt that because of my gender, this is not a field I can go into in high school so that was really, I was fortunate to have that experience.''
Thus, Ruby's high school experiences were such that she was able to talk to her teacher without any gender bias in the interactions and never felt discouraged to go into physics.

\subsubsection{Negative experiences in high school}
Four out of the 7 women described negative experiences during high school. Negative experiences include perceptions of and/or explicitly negative interactions with their peers and teachers.

When asked about her high school experiences, Chloe says:
``I think the competitive culture was quite toxic I'd say.''
When students applied to college and received early decision, she recalls,
``it became universally banned to talk about where you had gotten in because there is so much competition amongst students and so many people would feel bad [but] everybody knew where everybody had gotten in because it was the only topic of gossip our senior year.''
It seems like students in Chloe's senior class were fixated on their college status and students did not communicate with each other positively due to tensions related to college admissions. Chloe, who described herself as an introvert, also recalls negative experiences in her math class, in which,
``sometimes people would talk and when I would talk nobody would hear me, but when some of the quiet boys would talk, everybody would listen to them.''
She reflects upon and explains that her being ignored might be a result of students stereotyping her, e.g., male students assuming that she does not have anything substantial to contribute to conversations.

Karina feels guilty about going to a private high school, she says:
``If I think morally about private school education, I don't believe in it, so I felt guilty about having that opportunity. And it was a pretty wealthy school.'' Having access to better educational resources was a point of guilt and contention that she points out several times in her interview. She also shares that the culture of her high school was discouraging: ``I got kind of discouraged by it. It was kind of like a college prep type environment, where the sort of expectation was to go to an Ivy league school... I felt like everyone around me was thinking a lot about GPA and not really caring about the learning so I got a little discouraged...''.
Thus, in Karina's high school, the culture fostered high expectations for students to earn good grades to attend top universities as opposed to encouraging them to focus on learning, which made her feel discouraged.

Esme talks about her physics teacher and how she did not like the priorities he had for students:
``[My physics teacher] really didn't do a good job...We would just work on one problem, and he would just kind of vaguely explain the physics of it, but we mostly worked on trying to pass the AP [exam] rather than understand the concepts of physics. So, I didn't really like that class. It almost made me really quit physics, because...I didn't want to do just problems and not understand.''
Esme did not feel like she was learning physics, but rather working to pass the AP exam. Since the focus of the physics class was not to understand concepts, she did not like it and if it was not for a very positive NASA experience, she may not have chosen a physics and astronomy major in college.
She also mentions that her high school teacher called out only a few students who he thought would be the only ones to pass the AP exam if they try hard enough (implying that others were not capable).

When reflecting on how being a woman affected her interactions with her high school teacher, Esme says:
``but in terms of being a woman, my teacher was very strange, he... had a weird relationship with the girls in my class.''
She explains:
``he would always compliment them and...just be kind of unprofessional...''
She also recounts when her teacher would ask girls in the class for dating advice. 
In addition to a lack of emphasis on understanding concepts, this uncomfortable environment created by her high school physics teacher may have made it difficult to focus on learning.

Her interview also points to the fact that Esme had negative interactions with her male peers. For instance, she describes her peers as always thinking about ``the money" because they wanted to go into engineering and physics. She says that her peers were like, 
``{Yeah, I got this}, like they were very arrogant and they would kind of...mooch off of my work. They always tried to take my work...If we had partners, I usually did my portion [of] work, and I would try to make them do it, but sometimes they wouldn't and it would affect my grade so I would have to do everything.''
Esme's account of her male peers taking advantage of her work and leave her to do things on her own corresponds to findings from a previous study in which male students dominated group work \cite{doucette2020hermione}.

When asked why Esme found it difficult to communicate with her male peers, she says:
``It really seemed like they didn't respect me. I always thought maybe it was just my personality...I usually put up with things, but I have a limit and they would reach that limit many times.''
Despite the fact that she describes herself as having a reasonable amount of tolerance for her male peers' behavior, she thinks that her personality might be the reason why the interactions with them often led to friction. This is important to note as it may be a reflection of the masculine STEM culture and how it can make women think that they are the problem when it comes to negative gender dynamics and interactions with male peers.
She also noted that she made it a point to avoid her male peers. When asked about the gender distribution of her peers in her high school classes, she says that ``there were a little more men than women but I honestly can't say how many more because... I really would try to not interact with them unless they were someone I knew.''
This avoidance might have been stressful but it may have been a survival mechanism in that the avoidance may prevent her from experiencing any biases or microaggressions from her peers first hand.

Ruby says she liked her physics class and although she did well in the class, she says,
``I never felt like it was something that I could go into honestly before college. [At] my high school, no one ever talks about becoming a physics major and no one ever even tells you that that's necessarily an option, and until I got to college, I really didn't think that that was something I was even capable of if that makes sense.''
Although she liked her physics class and her high school teachers, she never felt encouraged to pursue physics as a college degree. Not having teachers encourage students to pursue physics could be extremely discouraging especially for students from traditionally marginalized groups. The lack of encouragement could also be interpreted as teachers not thinking students are capable of succeeding in a physics major, which may have contributed to Ruby feeling this way.

When recalling her high school physics teacher, Ruby says:
``even if my hand was up or if another girl's hand is [up]-that [my teacher] would always call on the male students before the female students.'' Recollecting that her high school teacher showed gender bias, she further says,
``I do think there was still this bias on his side of thinking, \textit{Oh, the guys are able to do this more than the girls. Well, oh, the girls will need your [more] help with laboratory experiments than their male counterparts.} And there definitely was that tone under the setting. But it wasn't obviously apparent.'' Ruby seemed to catch on to these biases which implicitly reflected that women may not be as good as men in physics or at physics lab. Despite the possibility that these biases may have had lasting experience on her,
Ruby says, ``I sort of just brushed it off because I didn't want [my high school teacher's gender bias] to get in the way of me doing something that I liked.'' It is great that she did not let a teacher's biased comments affect her self-efficacy or prevent her from pursuing physics but it could have affected other women in her class.

In terms of interacting with her peers, Ruby says:
``even my guy friends and my brother's friends would talk down to me when I had a question about math or physics and would definitely treat me like I was stupid or didn't know what I was doing without having any reason that we knew of for them to think that, and so I definitely felt scared to ask for help a lot of the time.'' Thus, it appears that Ruby may have faced some sort of biases from her male peers or her brother's friends but she persisted. She also explains that in addition to these negative interactions, her high school culture was competitive, ``It was a lot more of a competitive environment, at least, where I went to high school and, if you would go to someone for help, specifically if I would go to a guy friend for help, they would talk down to me while explaining the information. Well if I went to a girlfriend for help, they would be much more supportive and not at all condescending when helping with the information.'' In combination with these poor interactions with male peers, a competitive culture may have contributed to her initially being afraid to major in physics during college.

\subsection{Interactions with college peers}
In this section, we describe the experiences the women had with their peers in college physics courses. These experiences include positive and negative interactions and perceptions.

\subsubsection{Positive interactions and perceptions}
Only a few women described any positive interactions with and positive perceptions of their male peers. One possible reason why there were not that many women who mentioned positive experiences with their peers may be due to the fact that there were many more negative interactions that stood out to them.

Chloe explicitly says that her interactions with her peers have been good and that she does not think that her gender excludes her in spaces. She clarifies that she has always been in classes with other girls, so this could have been a possible reason for her feeling included. With regard to forming groups, Chloe explains,
``I have made groups for all my study groups and people I really talk to are other girls and that has been the most comfortable for me.''
It may be that Chloe has surrounded herself with supportive people, specifically women who tend to be more approachable, thus avoiding any negative interactions and experiences with male peers. For instance, her study groups consist of other women.

Ruby is the only one who explicitly mentions that men and women work together well. In particular, she mentions not worrying about asking male peers for help and how she perceives her interactions with them to be positive:
``I think [my college] has a very good environment in terms of men and women working together and collaborating...I really haven't felt that fear of asking a guy for help versus asking a girl for help here as much as in high school.''
Despite these positive reflections, we note that she is speaking about her college experiences relative to her high school interactions.
In the next section, we will discuss negative experiences Ruby had in college with male peers.

\subsubsection{Negative interactions and perceptions}
All of the women recalled negative experiences with their male peers during college. At the end of this section, we include Paulina's experiences as separate from the others, but still under this theme. The reason for doing so is that her experiences are different due to her specifically pointing to both her race and gender as being aspects of her identity that led to her negative experiences with both men and women students.

Cathy did not provide many experiences about her peers in physics but describes that she has had many negative experiences with peers in her computer science courses. 
According to Cathy, peers in her computer science courses are intimidating to approach to ask for help since it seems like they have a lot of background in computer programming which is very relevant in the courses. She also notes that for computer science courses, ``a lot of people come in with a lot of experience,'' also making it difficult to reach out to peers. Without any actions from the instructor to put students in groups in which they can learn from each other on projects, the structure of these courses did not support peer collaboration, which made it more difficult to connect with other students.

Chloe shares an experience in which she had an exam and following the exam, she heard some students talk negatively about it. For example, they did not like the format of the exam. She says that she
``felt bad vibes from the conversation, and most of them were boys.'' Chloe's bad vibes about male students complaining about the exam in a course taught by a woman instructor reflects
an uncomfortable environment.
This may also illustrate how men may be more critical of assessments including their format, especially if it was given by a woman instructor.

In regard to working in groups, Chloe says that she does not feel as comfortable working with men as she does working with women. She also reflects on why this may be the case:
``well, I guess it's all just an inherent thing that's still part of our culture... I know that's a difficult problem to solve.''
According to Chloe, she mostly interacts with women peers within study groups and feels comfortable doing so. When it comes to working with male peers, she says, ``I'd only really feel comfortable working with the boys if I'm already friends with them outside of the classroom, like they're one of my closer friends.''
It seems like for men, there's a higher threshold to overcome in order for them to be deemed comfortable enough for Chloe to work with them.

When asked about who is part of her study group, Francis explains that she mostly works on homework with women because of the negative experiences she has had with male peers:
``in one of my physics classes, if me and my partner had different answers, it kind of felt like he'd automatically assume that he was right, which wasn't always the case, and to me that just felt like a very male thing to [do]- So I guess, I generally would select to study with a woman.''
Thus, having negative experiences with men in her study groups has affected whom she currently chooses to work with. She also attributes this behavior of automatically assuming they have the correct answer as a ``very male thing to do", which hints to a masculine physics culture.

She also says that men have more confidence in answering questions and their confidence takes up space in the classroom, saying that it felt like ``there was a lot more men in the class than there actually are''. In contrast, she does the opposite:
``I wouldn't answer a question unless I know for sure that I'm right...''
This correlates with prior research in which men in physics courses are more confident in answering questions \cite{santana2023effects, liimpact2023}.
When asked if she thinks that these students could be showing off, she says,
``I remember one guy who... would ask questions that kind of felt like he knew the answers [to]. He just wanted to show that he was thinking about this one specific thing.''
This type of report of male students showing off reflects the physics culture in which men feel comfortable doing so. Regardless of whether this was intentional or not, women tend to perceive this behavior as intimidating or these types of behaviors can reduce their sense of belonging.

When asked about what the vibes were in the classroom and if there was a dominant group, Karina recalls that men were more dominant and they participated more.
As a result, she says,
``they're...sort of like blocking out the space to ask more basic questions to understand the material at the most basic level.''
She further provides a similar insight as Francis did: 
``there would be men asking like questions that perhaps are built to make himself seem really smart and so it's suffocating this space to ask sort of more basic level questions for other people.''
Again, Karina adds to Francis's point of men often showing off and wanting to be perceived as smart by asking certain type of questions, by saying that they are ``suffocating the space" for other people to ask more basic questions. From her phrasing, we interpreted these types of behaviors to be very negative experiences.
These behaviors can also alienate and make other students anxious by creating overly high expectations for types of questions that can be asked or knowledge that can be brought to the table during a physics class.

Karina shares that she has a different approach to learning, she takes longer to learn the material. She also explains how this negatively affects her interactions with her male peers:
``in labs I sometimes felt like the group was moving on... and I didn't want to speak up...maybe just because of pride or something. Also, I didn't want to be like `Oh, hold on. I don't understand what's going on here.', `Can I try this?', `Can I do this?' and I'm not sure how much of that has to do with gender versus, you know, just the type of learner I am and the way I work.''
While on the one hand, Karina seems to be contemplating as to whether her gender was a reason for these negative interactions and the fact that she did not feel comfortable speaking up about letting her try certain things and slowing down, she also puts the onus on herself and contemplates whether it is her learning style that is making it difficult for her to collaborate with her male peers in the physics labs and giving her inadequate opportunities to delve deeper into the concepts and be on the same page with them.

Although Ruby shared previously that she perceives the collaborative environment to be positive, she also shares that her interactions with her male peers are varied. For example, many of the men in her class, she describes as helpful, ``willing to help'', and do not ``approach things in a condescending manner.'' However, she also identifies some men 
``who act like a know-it-all.''
She also attempts to reason why they exhibit this behavior:
``I don't know if they do it because I'm a woman, but sometimes it feels like that, where they're like mansplaining so often in and outside the classroom, there is definitely a culture of mansplaining.''
For example, she recalls working with a male peer lab partner via Zoom. She says that her partner was extremely condescending and, ``...I think he just thought he was smarter than everyone else in the room at all times and always had to be the one to answer the question first, always had to be the one to know it...I ended up switching my partners actually because I just could not deal with that.''

While on Zoom, she says that her partner would dominate breakout room conversations and assignments, leaving her and her other woman partner behind:
``He would always just solve it himself and then present the answer, without even giving us a chance to try and that really bothered me because it made me feel like...he thinks oh, because he knows it, that all of us cannot be smart enough to even try...and it made me feel like I was being talked down to from him, thinking that I was less smart than he was.'' The fact that this happened over Zoom may have made this situation especially difficult to be in. Ruby also adds that her partner would not collaborate on the written report and would write it himself. She says, ``he would act like I didn't know anything that I was doing, when I was fully capable of doing it myself and that really made me feel...angry about it because... I love collaboration, I think it's so important, and I really liked when I'm able to work with somebody, we're able to combine our own knowledge and figure out new stuff, and with him, I felt that I was limiting my own knowledge, because I wasn't getting access to even try. And I felt he was mansplaining things all the freaking time. I really did not like it.''

When asked if she thought that this behavior was specifically towards women, Ruby says, ``100\% yes.'' Again, she explains that she thought that the student thought that he was smarter than others,
``but he also believed that he was helping in a way. He thought that, \textit{Oh, these women can't do it themselves, I'll help them, I'll do it for them}... But in my mind at least, I believe that's why he did, that I think he thought of women in a sense, that they were less capable and more vulnerable than men, and so he wanted to be this savior.'' Thus, in her experience, this male peer thought that he was helping women by answering for them and doing all the work, when in reality, it had the opposite effect.

Cathy recalls her physics REU experiences at large research universities, some of them being very negative. In one of her REUs, she was the only woman in a group of nine students and she was not familiar with the research topic. As a result, she felt like she could not ask questions or reach out to others, thus making her feel excluded.
She explains that:
``everybody else knew so much more than me and they kind of talked down to me, sometimes it was just really overwhelming and intimidating and it definitely harmed my confidence a lot, because since I didn't feel like I could ask questions, I really didn't understand what was going on, and I got super behind, which made it so it was harder to ask questions, it's just kind of the cycle and I don't really feel like I learned anything in the end because I was just so overwhelmed the whole time.'' Without any support from REU mentors, the fact that the other male students gave her the impression that they were more knowledgeable than her and they talked down to her made her intimidated to ask questions, potentially making her not benefit from the entire REU experience when a supporting environment would have made it very beneficial. 

\subsubsection{Paulina as an island among her college peers}

In this section, we present Paulina's negative experiences with her college peers. The reason for separating her experiences with peers is two-fold. One reason is that she repeatedly expressed how both her gender and race affected her interactions with peers. Secondly, she expressed how isolating these experiences were, so much so that they resulted in her becoming an island and not interacting with anyone because of these negative experiences. We also note that Paulina shared the most negative experiences compared to the rest of the women.

\textbf{Being perceived negatively by peers}: Paulina is very aware of how the intersection of her marginalized identities (i.e., gender and race) impacts her in the physics learning environments and how she may be perceived by her peers. In the interviews, other women shared stories about how their identity as women affect how male peers perceive them or interact with them. However, Paulina explicitly described how her multiple marginalized identities of being a woman of color interact and make her negative interactions with her college peers at a predominantly White institution (PWI) more complex in nature.

She says that being a woman of color is complicated because,
``people perceive me as being different in the classroom because I'm a person of color and also because I'm a woman.''
She says these visible identities are apparent and 
``that's...the first thing they see and it's how they assess my value or what I can contribute to the classroom.'' As one of the few people of color in her program, we note that this may be extremely difficult to navigate as she constantly thinks about how others are perceiving her. She also says that students simultaneously are judging her and assessing her intellect and the value of interacting with her and what she will bring to collaboration of physics problem solving. Similar issues have been documented in previous studies involving interviews with six physics graduate students who were Black women \cite{rosa2016obstacles}.

Paulina explains that she even feels judged amongst other women in her course,
``there's also been this expectation that I am just not up to par with them, because [of] my racial background and so I feel there's some intersections in terms of my experiences in STEM with women...I definitely know that beyond the realm of just gender, race also plays a really big role in it.''
In Paulina's experience, she doesn't necessarily find solidarity amongst other White women either because she is a woman of color. Since her university is both small and a PWI, finding support and solidarity in other students who look like her can be extremely difficult.

She adds that whether she interacts with students inside or outside of the classroom:
``there is more of a willingness to either poke fun or underestimate me because of the intersectionality of my gender and race.''
She believes that the interaction between her multiple marginalized identities in physics has led to her negative interactions with her peers.
We also theorize that her comment about the other students ``poking fun" of her because of an intersection of marginalized identities may be due to students stereotyping her. These marginalized identities have resulted in Paulina feeling like an island among her peers and she interprets the negative interactions with them as at least partly due to them feeling that she does not have much to contribute in a collaborative learning environments due to her marginalized identities.

\textbf{Being ignored by peers}: Much of Paulina's experiences are centered around her working alone on assignments because she is ignored by her peers during group work.
Below, we discuss how her intersecting marginalized identities negatively impact her throughout and play a central role in her interactions with peers.
Paulina explains that her peers will not value or listen to her contribution in groups because they make assumptions based on her identity as a woman of color:
``I think there's this assumption that... because I am a woman of color in STEM, I don't necessarily have the same background or knowledge as they [do], so a lot of the times I'll bring [up] points and it'll just be ignored and that's something that I've had to deal with a lot.''
From this account, it is clear that Paulina is aware of stereotypes her peers may believe in, which ultimately result in them not engaging with her appropriately during group work.

She reveals that this semester for the first time, she has been placed in a group with students who she has worked with before in her class. These students have similar physics backgrounds to Paulina and she feels that although there seems to be some effort to meet and work through assignments, it is difficult. She reflects upon working with these students outside of class that she has interacted with in class earlier,
``[t]hey either stick to doing things on their own, or like they already have a background with other students in the class, so they'll just reach out to them, and if we happen to be in the same group, they'll usually just choose to staying engaged with the students they already know, rather than work together as a group with me who they might not be as familiar with.''
Thus, according to Paulina, other students feel more comfortable with others who share their identity and whom they know more personally or have interacted with more. The fact that these students have all been placed in a group with Paulina to work outside of the class and she is still feeling excluded (both inside and outside of the classroom) is very concerning.
This formation of a clique makes it difficult for Paulina to meaningfully collaborate with other peers (including other women) as they choose to talk to other students in other groups who they do know and have similar identities (e.g., gender and race), as opposed to her, who is placed in their group.

\textbf{Working alone:} Because Paulina is so isolated and feels negatively perceived due to her multiple marginalized identities, she usually resorts to working on her own, which also takes up mental energy and extra resources, e.g., time.

Paulina explains that although some of her classes incorporate group work, there is no effort to communicate or work on problems outside of class. Even for the case mentioned earlier in which she has been placed in a group for studying outside the class, she feels that it does not work effectively, and she is left to work on her own. She says that as a result:
``no one is reaching out to each other to talk and work things out together so by default I just feel like I have to end up doing things on my own.''
Thus, even when she has been placed in a group to work outside of class, her own group does not communicate about when they should get together to study or do homework, so she resorts to working 
alone. It is unclear whether students who have been placed in Paulina's group are meeting with other students with similar identity outside of the class to study and do homework. In particular, she is not aware of what is happening since nothing happens through the messaging system they were supposed to communicate through when they were placed in the same group for out of class study.
Whether this lack of communication is directed towards Paulina because of her identity or caused by other reasons such as none of the students in the group she has been placed meeting regularly, it is interpreted as lack of community by her.
Thus, she is unable to lean on her peers as resources and collaborators, and can only rely on herself.

Inside the classroom, when she is ignored by her peers, she feels like she cannot rely on her group mates either. She feels like her only option is to reach out to her instructor and says she feels like,
``I was expected to learn things on my own and if I didn't verify [that my problem solving process was correct with the professor] it reflected badly on my own personal knowledge and understanding of the subject.''
Throughout her interview, it is clear that Paulina is not only aware of how others negatively perceive her and exclude her, but it also appears to take a toll on her. For example, she feels like she has to overachieve and 
``...prove that I am just as knowledgeable as everyone else in the space.''
Paulina's experience is also consistent with other women of color doing more than just learning physics \cite{santana2022investigating}.
The fact that she feels that she has to overachieve and prove that she is knowledgeable like her peers is an additional burden on her because she does not want to confirm any stereotype others may have related to her identity \cite{steele2010stereotypes}.

Because other students would not work with her even in class even though she was in a group with them,
she could not complete in-class assignments during class like her peers did. She says that she would fall behind and be ``left to my own devices'' to work on problems. She also describes a cycle where she could not use her peers as a resource, i.e., ask them questions, and would have to finish the assignment outside of class. On the other hand, her peers would be able to complete assignments during class.
Not only does she have to manage feelings of being excluded and negatively stereotyped by her peers, she also has to spend extra time outside of class to finish her class assignments.
We can see how her multiple marginalized identities in physics can take up emotional and mental space that could be used for learning physics and problem solving.
She has to use additional time and resources to complete her in-class assignments due to the exclusion by peers and navigate her physics journey alone.

Paulina was the only student to mention the weekly meeting of undergraduate physics majors in the student lounge, which is a place designed with the intention to establish a community culture. She says that these are informal meetings where students can connect. However, if she misses a single meeting, she feels completely left out:
``if you end up not going to one of these meetings one week, for example, it almost feels like you've fallen behind...you lost so much time where when you do come back, everything feels different and it feels-it's difficult to kind of get back to a position where you feel familiar enough with everyone's engaging actively...it just feels like everything is very fast pace and almost divided into different groups where, if you don't attend, then you're immediately out of the loop.''
This is especially concerning due to how small the department is and how essential it is to create a space for all students regardless of their multiple marginalized identities to connect and feel safe. Paulina already feels isolated due to her race and gender in her classes. Feeling excluded in informal settings is concerning because they are supposed to be low-stakes and more casual and could provide opportunity for her to bond with other students.

\subsection{Interactions with college instructors}
In this section, we describe the experiences the women had with their college physics instructors. These experiences include Positive and Negative interactions and perceptions.

\subsubsection{Positive interactions and perceptions}
Most of the women explicitly said that they have had good experiences with their physics instructors. We note that Paulina was the only woman who did not say this in her interview.

When asked about the gender dynamics in her courses, Esme brings up her academic advisor, who she had positive experiences with. Esme says that she has become very comfortable with her advisor because of her role both as her advisor and instructor. Esme also explains that during class, her advisor
``wants us to be comfortable with the people that we interact with. I know we're about to do a project and she's trying to make sure not to put the women in our class as the only woman in a group, I know she's working on that and I definitely am grateful for that, because I definitely would not want to be in a group by myself as a woman with men...''
She notes that at least her advisor puts in effort to make sure students feel comfortable with peers that they interact with.
It also appears that this instructor is aware of the negative impacts of a woman being the only woman in a group \cite{heller1992teaching}.

When asked about her overall college experience, Ruby feels 
``very fortunate to go to school at [my college] where the instructors really want you to succeed and I felt so supported in physics, math, and honestly all of my classes here.'' 
She feels like her instructors are encouraging and supporting her academically. This is a good sign that Ruby has positive experiences with her instructors. However, this does not specifically mean that she only has had good experiences with all instructors.

Both Ruby and Francis mention positive interactions with their instructors during office hours. Ruby explains that she had difficulty grasping certain concepts in her courses,
but whenever she talks to her professors during office hours, they are very helpful.
In Ruby's experience, her professor normalized struggle and helped her realize that not all students must understand a physics concept at the same rate. She also says that she feels comfortable to ask professors questions during office hours and that it is a welcoming environment.
Francis explains that she does not attend office hours, unless she has a specific question. However, when she has gone, she has had positive experiences. She says:
``it was definitely helpful.''
It appears that this particular professor provides good support during office hours and did a good job at answering her questions whenever she went.

Francis also explains that one of her favorite class period was one that focused on marginalized students' identities in physics and their experiences in physics. In this class, students read articles from marginalized people and then they discussed them during class. The discussions focused on how physics needs to change to become more inclusive. Francis says,
``including something like that at some point in your undergraduate studies is definitely a good thing, trying to make sure everyone is aware and actively thinking about how you can do better to make things more inclusive and I guess, peers just acknowledging that...people are coming from different backgrounds and not assuming that you know more than someone before talking to them.''
Francis praises this instructor for incorporating such a discussion into a class and the value of such discussions for all students.

Chloe says that overall, her experience in physics has been good and that the physics community is ``welcoming and positive.'' She also adds that her instructors
``have really made attempts to make their classroom environments better.''
She recalls an example of one of her instructors trying to create an inclusive learning environment by asking a trained student to take notes on the instructors' behavior and classroom dynamics. This trained student also talked with the students in the class and asked questions about the classroom environment and how the students feel. Then, they shared this information with the instructor.
It seems like this specific instructor cares about how their students are interacting with each other and how they engage with their students. This practice is very useful as it can help instructors understand how students experience their physics classroom environment.

\subsubsection{Negative interactions and perceptions}

Here we share negative interactions with and perceptions of some of the instructors that the women had. For some women, their negative perceptions stem from lack of action or interventions during negative interactions with peers. This, in particular, is a reflection of the lack of disciplinary action (if male peers' behavior is not appropriate) or shows lack of efforts in general to create an equitable and inclusive learning environment by instructors.

As part of the interview, when asked if her professors can do more to prevent male students from negatively stereotyping women in physics, Chloe says: ``I'm not sure if they noticed it. I think the professors who tried to notice it do [see biases], but if a professor isn't really trying to actively improve equity in the classroom then it's easy to miss.'' To her, it is not apparent that some of the instructors are aware of biases that male students have towards women. According to Chloe, in order to notice these things, instructors have to be, e.g., alert to male students dominating or showing off or de-valuing women and actively improving their classroom environments. This awareness is essential to instructors then making changes to improve the learning environments.

Ruby says that while working with her lab partner who was condescending towards her, one of her instructors intervened and one of them did not. She says, for one class,
``The instructor for my class actually did notice that the student was being very difficult to work with and that he wasn't collaborating with the group, and she I think spoke to him about it.'' Although this lecture instructor addressed the student's behavior, the lab instructor did not. She explains:
``But in the lab I don't know why they didn't notice, I think part of it was, a lot of our experiments were done outside of the physical classroom because, again with COVID, it was hard to sort of build those relationships with professors...'' Ruby also explains that when students are submitting assignments on time, these issues might be overlooked:
``a lot of professors don't go out of their way to look at the problems and a lot of professors, when necessary, will assume [things are fine] or [don't] want to have to deal with anything like that.'' 
Another reason why the lab instructor did not address the student was
``because [my lab partner] knew the answers to a lot of things and...he is smart, he's just horrible to work with and because of that, it's possible that, that gave him the right... to act how he was acting.''
Thus, Ruby suggests that because her male peer appeared smart, that may have justified his behavior to the professor. Thus, Ruby reveals that instructors have a role in addressing students' behaviors and not being biased by whether the student appears to be smart because these types of student behaviors are tied to the physics culture.

Paulina shares a few experiences when her professor singled her out and assumed that she was falling behind when she was not. She says,
``the instructor has kind of singled me out in the classroom and asked me questions continuously trying to I guess make sure that I'm keeping up with the class, even though there's been no signs of me struggling with the material.''
It is possible that the professor is aware of Paulina's marginalized identity as a woman of color and overcompensated by asking her questions in a way that she felt stereotyped. In Paulina's view, this came across as the professor underestimating her, as she also perceives her peers to do.
We note that it could be that this instructor had good intentions. However, the physics culture was not equitable and inclusive so it was not possible for Paulina to view the instructor's intentions and actions as positive, and she perceived this experience as the instructor singling her out as opposed to checking-in on her.

When asked if she's been in any situations where she had to stand up for herself, Paulina shares experiences in which other students in her classes stood up to the professor:
``someone else has stepped up and [was] like,... \textit{I feel like I'm not being accurately listened to, or represented in this atmosphere}.''
She also explains that as a result, physics instructors will be surprised and during the next class would address students, saying,
``\textit{Hey, we should be engaging with each other} or trying to discretely say...there should be more interactions with individuals that aren't necessarily White or that are women in STEM...'''
However, Paulina emphasizes that despite these types of occurrences, the overall feeling is that:
``a few weeks later, it kind of goes back to what it used to be.''
According to Paulina, when put in these types of situations in which students call out professors directly, although a professor may have superficially tried to address the issues brought up by students, it is not very impactful or long-lasting. We suspect that the physics culture is so established that when a professor addresses an issue superficially and passively, it does not have any long term effect on the culture since it was not 
reinforced in any way and things go back to the way they were very soon.

\subsection{Reflections on how various aspects of their identities affect physics experiences}

In this section, we provide reflections from women who share how their specific identities negatively affect their experiences in physics. These reflections are helpful in assessing their awareness of how their specific identities affect their experiences in physics and contemplating strategies to ensure that these identities do not negatively impact how they navigate their path in their pursuit of their physics degree.
We note that some of the women had difficulty naming their gender as one of the reasons for why they may have had negative experiences with male peers, or did not mention gender bias to explain why their male peers were condescending even though they described very well going through these experiences.

Chloe describes herself as introverted and expressed how this part of her personality has made it difficult to navigate group work. She notices that physics is a very social environment saying:
``I think personality is the thing that isn't always talked about like being very social...versus somebody who's not...most physicists are nerdy and I tend to be more introverted...outside of the classroom. I think personality does create some divides...''
Thus, here Chloe shares her idea of a physicist being nerdy and her not relating to these types of identities.
She suggests that for someone who is more quiet like her, it may be difficult to try to engage in conversations with social and extroverted physicists.

We note that several women reflected on how their identity as women impacted their experiences in physics. Some women felt that the lack of women in physics spaces negatively impacted them.

Francis was one of the few women who felt she did belong in physics. This could be attributed to her high school experiences as she went to an all-girls high school in the UK. With that in mind, Francis explains that sometimes when there are very few women in a class or as the only woman in a group. She says that despite feeling like she belongs, 
``...it was just weird because that wasn't really something I've experienced before being the only woman in a space. I'm not really the kind of person who speaks up a huge amount in class anyway, but yeah I didn't really know what was weird about it's just something different.''
She does not explicitly say that she feels alienated if there are very few women or she is the only woman in a group with a bunch of men participating.
We hypothesize that her being quiet and being from a different country might make her more accepting of different norms she is being exposed to in her classes even though she finds the dominant participation of her male peers  ``weird" and feels this is a ``different" environment compared to what she was used to in her all-girls high school. It is even possible that her male peers were not as condescending to her as they were to some of the other women due to her being from the UK.

Karina explains that sometimes she feels pressure to ``be smart,'' particularly in her classes in which there are not that many women.
She says it feels like
``...you're representing women.''
Whether or not she is consciously aware that this pressure is partly from the lack of women's representation in her classes and stereotypes about who can excel in physics, Karina feels it.
Here Karina may be alluding to stereotype threat, the fear of confirming a negative stereotype about oneself, i.e., women are not capable of doing or excelling in physics.
She may not be fully aware that she is representing women in physics, but she still wants to appear smart and knowledgeable in her physics classes.
This may also hinder her learning in that this awareness and resulting pressure takes up cognitive resources that could have been used to learn physics.

Karina confesses she is afraid of not appearing smart. Again, the fear and anxiety may be correlated with stereotype threat.
We also see Karina 
reflect upon these types of thought processes:
``Maybe logically I wouldn't think that those people would think women are bad at physics, if I asked them to slow down, so maybe if anything, that would be like a subconscious fear because logically that doesn't really add up to me, but I think subconsciously maybe, yeah.''
Although Karina is confident about her own ability, she seems conflicted about not appearing smart in front of her peers:
``I think I actually have a lot of confidence in myself and my ability...I don't think I am too worried about sounding dumb, yeah, which [may sound like] I was saying kind of the opposite before.''
She admits that it may appear that she is contradicting her previous statement about being worried about representing women in a positive way.
Karina is an example of how conflicted a woman physics student may be about how their identity may shape their experiences.

Karina explains that her privileged identity, i.e., having private education, has cushioned her from feeling isolated. This is particularly interesting because she explains that the further she progressed in her degree, the less women there were. As a woman, this may be challenging to navigate. However, Karina explains that the further she progressed in college:
``I was more comfortable in the department, just like I knew all the professors. I knew the space. I had been you know living in [college town] at [college] for longer by the time I was a junior/senior, so I think I didn't so much feel like isolated or alienated because of that, maybe I would have felt that more, earlier on...I think again, I've been pretty cushioned in physics, I haven't really had to face a whole lot of awful things...''
Thus, Karina clarifies that her negative feelings, e.g., the pressure to appear smart and represent women in physics, stemmed from not seeing many women in physics and she didn't particularly have direct negative experiences with her peers based on her identity as a woman. We also see that her familiarity with professors and the physics space helped her navigate these physics experience.

When asked about if/how her race/ethnicity played a role in her experiences in physics and astronomy, Esme explains that she was one of two Hispanic students in physics at this college and one of two women during her experience at NASA. She reflects on this issue saying, 
``I think more gender [than race] in my experience has been affected.''
This reflection seems consistent with her previous experiences that she describes, in which gender was more salient in her experiences related to physics. Again we hypothesize that this may partly be the case because of her visual characteristics not ``manifesting" her ethnicity in the same manner as Paulina.

We note that Paulina constantly emphasizes how her multiple marginalized identities as a woman and person of color were salient in her interactions with peers and even instructors. She attempted to disentangle the two in the context of her interactions with her women peers, i.e., she felt that they judged her differently based on her racial identity (hypothesized by Paulina). However, throughout her interview, both identities were constantly brought up by her when explaining her interactions with others in the physics learning environments in college.

Paulina also mentions that sometimes she works with two students, one being a man of color. She says that they usually consult each other for help when it comes to problem solving. Although she has not had any conversations with him about what his experiences are like in physics, being a man of color, she believes it might be easier for him to seek out help from other students in the classroom. She says that he might be more comfortable approaching other students if he needs help. Paulina also speculates that
``it becomes a bit easier for him to engage in the classroom with other individuals...it feels as if his first response is, [to] select, seek out other people of color in the classroom and have...conversations about it [with other students].''
Although they do work together, she thinks that his identity as a man might be a privilege over her multiple marginalized identities, as she has had difficulty getting other peers to take what she says seriously.

\subsection{Supports and Role Models}
In this section, we share who or what has supported and sustained the women and kept them on their path during their undergraduate physics journey. This support may include role models, i.e., people who they looked up to. 

\subsubsection{People providing academic support}
In the previous themes, such as Family providing academic support, we discussed how some of the women already highlighted how their family provided support by helping them with their homework. For many of the women, their family members were also role models, especially if they had careers in science. For example, Cathy shares that her sister and her parents were some of her role models and provided foundations and inspiration for going into a science major. Many women noted that their professors were their role models. Some of these professors were also their academic or research advisors, thus serving multiple roles and having multiple contexts for providing support.

Another important source of support Chloe mentions is mentoring. The mentoring program she describes is part of a gender inclusive STEM club where upperclassmen partner with lowerclassmen. Chloe specifies that mentoring program is all girls and the upperclassmen help the lowerclassmen with their classes. She says that her mentor was:
``really helpful and show[ed] me that I can apply for these research and these internships and to pursue these opportunities, yeah and just seeing people who are older and are currently doing what I am aspiring to do in the future is very inspiring.''
This mentoring program connected her with more experienced students and particularly her mentor who guided her through opportunities such as research programs and internships.

\subsubsection{People providing emotional support}

Chloe shares that one of her professors is really inspiring because of a shared identity. Chloe considers herself to be an introvert and her professor is also quiet, so she feels inspired by her because:
``she is kind of a quieter person and most of the professors that I've met, you ask them a question and then they talk very confidently for hours and she's not like that...so meeting somebody who's not like that but is still very successful in the field, yeah, that is very inspiring.''
This type of role model is valuable as the stereotypical image of a physicist is usually either social and outgoing (like those portrayed in the media) or awkward and standoffish. Seeing someone who shares similar characteristics is helpful for someone like Chloe, who does not have the stereotypical personality of a physicist, to see herself as capable of being successful as one.

Paulina also shares that her role model is her academic advisor, who is also her research advisor and someone she had classes with. She says that he encouraged her to pursue physics as a college major:
``he was one of the first people that kind of sat down with me and kind of listened to me speak about my passion for physics and he was like \textit{Hey it sounds like there's something beyond a hobby}, \textit{I think it would be very interesting for you to take more physics classes at a high level and see how you feel about them}, and having someone that took the time to have a conversation with me and help me figure things out was really meaningful to me.''
Another important factor, she adds is that he shares a common identity of being Latinx:
``He's one of the few other Latinx professors that I've seen on campus specifically in the physics department so it's very, it's become very meaningful to see someone from a similar background as me kind of succeed and serve as a person that I can go to and he's very encouraging.''
As someone who attends a PWI and whose peers are predominantly White, seeing and connecting with a faculty member who has a similar identity is very important to Paulina.

Ruby shares that one of her role models is her academic advisor because she helped Ruby consider majoring in physics. Following that conversation, Ruby also turned to her family for support/advice:
``They'd all told me, \textit{You gotta go with your gut}, \textit{you have to trust your gut because you know what you can do and you can do more than what you think you can do}, and that really stuck with me. This idea of you're gonna put limitations on what you think you can do, but you can always go past that limit a little bit, even if it's just the tiniest bit you can push past that and I look up to my brothers a lot so after hearing that from them and talking to my parents as well, I think I was able to let go of that fear, because how much I wanted it [pursue physics] and how much I realized that fear is so in our heads.''
Here Ruby was able to go to her family as a soundboard for advice and reassurance when she was worried about the hurdles she will encounter majoring in physics after her academic advisor suggested she consider majoring in it.

\subsection{Suggestions to improve experiences}
In this section, we present the suggestions women provided to help improve their experiences. We separated these suggestions into two categories, the first is personal advice that has helped the women persist in physics and the second is suggestions for instructors they think would help improve their experiences.

\subsubsection{Personal Advice: The "Band-Aid" Approach}
We note that although this personal advice may have helped a few of the women, it may not be useful for others. We also labeled this personal advice as \textit{The "Band-Aid" Approach"} because although this may be genuine advice from the women in physics, it only scratches the surface of the issues that traditionally marginalized students face within physics environments. In particular, this advice is more oriented towards personal action as opposed to more systemic change in the physics culture, which is discussed in a later sub-theme.

Cathy has served on an undergraduate committee for physics majors in her program. She says that this opportunity has been beneficial for her because
``...you can get involved in making whatever changes you feel would be helpful and having the department listen to the students...''
Through her role on this committee, she was also able to host events for her peers, such as a physics GRE program and study program. She also emphasizes that:
``I liked that once I was involved, I could then kind of make the events that I thought would be helpful so I'm glad I did that.''
It appears that this may be personal advice based specifically upon the experience with undergraduate major committee that helped Cathy.  We note that not every student may have this opportunity or may feel safe being in these types of roles. However, being involved in different types of departmental committees and activities based upon interest may be valuable for students.

Ruby provides some advice that has helped her persist, which includes having a growth mindset. By adopting a growth mindset, she learned that the focus is more so on the process:
``it's about just practicing and trying and also just learning as much as you can. And I think that advice has really helped me in that feeling support, even when I'm not going to be even close to the smartest person in that room. [It] has really helped me.''
Having growth mindset may in fact be useful for students to adopt. However, it is especially effective when instructors and the physics culture encourages and reflects a growth mindset as well so that the onus is not only on the student, like Ruby, to adopt a growth mindset in order to excel in physics. In particular, when instructors have growth mindset about their students' potential and believe that all students can do well in physics, it can help students develop growth mindset and believe that they can become good at physics by working hard, working smart and taking advantage of resources.

Francis shares that she thinks that believing in yourself would help future women in physics. She also encourages others to ignore people if they act like know-it-alls:
``just because other people, especially men, are talking more and acting like they know more, it's not necessarily true and you know, [you] are just as intelligent as them.''
Thus, Francis' advice based upon her personal experiences is to ``believe in yourself" and although this is a good thing to do, it may be difficult for women in physics if they do not have the support and it does not appear to them that other people believe in them (as in the case of Paulina). This advice may be viewed by some as putting onus on the student to change to fit in to survive and make it through the physics culture. Having instructors believe in their students is necessary for traditionally marginalized students to truly believe in themselves with regard to excelling in physics. These positive changes can be heralded with changing the actual culture of physics that can be usually intimidating for many students.

\subsubsection{Advice for instructors}

In this section, we summarize the suggestions for instructors provided by these undergraduate women. We have bolded their suggestions. However, we note that many of their suggestions overlapped and would fit into several general suggestion categories.

\textbf{Avoid forming opinions about students based on background}:
Karina shares some insight on how instructors should not make assumptions about students based on their background. This can be problematic because
``...especially [at] a small school like [my college], where you might have professors for multiple years, you might come back to that professor that you've had your first semester and I think that's important, that's something that's important for professors to know.'' 
Besides potentially having the same instructor for multiple courses, these instructors also serve as research advisors and mentors, so it is especially important to have good perceptions of (and relationships with) students. In addition, having preconceived notions about students may unintentionally lead to instructors stereotyping students or making generalizations, particularly about their ability.

\textbf{Address and dismantle biases against women}:
Ruby shares that it is important for instructors to explicitly communicate in their actions and words that women are just as capable as men in terms of their physics ability. She says that instructors should
``mak[e] it a very, very clear fact that...men are no better at physics than women. Just because there are more men in the history of physics does not mean that men in your classroom are going to be better at it than you are because of your gender, because that has no influence on how capable you are [of] learning the subject, even if people tell you otherwise.''
Changing the physics culture to reflect beliefs about all students as equally capable of excelling in physics is key. Having instructors emphasize this and continue to do so not only through verbal affirmations but via their actions as well would be necessary to reflect how deeply they support this tenet.

In one of her reflections regarding her gender identity, she provides a suggestion for instructors to address biases against women.
For context, she emphasizes that 
she has to work hard to prove that she is capable:
``I feel like as a woman, I have to prove myself for people to look at me and go, \textit{Oh, you're smart and dedicated and you like science}, I have to prove that first. It's not implicitly assumed and nobody assumes that when they first meet me... for many men in my classes it's the opposite, where it's like you initially assume that this person is really smart and gets really good grades just because they're male in science.'' She believes increasing the number of women in physics programs could help reduce these biases:
``if many other girls are also doing really well, I think it's a self perpetuating cycle that as more women get involved in science, the more everybody... realizes that we're on the same playing field.''
So having instructors dismantle these stereotypes and biases about who belongs in physics and can excel in it through their actions can be very impactful.

\textbf{Reassure about lack of representation of women in physics}:
Ruby shares that she is aware of the historical lack of representation of women in physics but she would appreciate her department to acknowledge that while uplifting women:
``I haven't really ever been told ...\textit{just because there aren't a lot of women that did it before you, doesn't mean you can't do it}, but it'd be so reassuring and helpful to hear...I think that that would be beneficial to sort of creating that comfortable environment and creating that confidence.''
Thus, Ruby emphasizes that acknowledging this part of physics history would not only educate departmental members but also students would be aware of this going forward in their major. Following this acknowledgement, reassurance would be helpful so that women are not worried about majoring in physics and the department feels supportive to them.

\textbf{Create a better representation of women in physics}:
Esme shares that having more representation of women in physics would be beneficial:
``seeing more women be the forefront, be the face of physics, because usually when we think of physics, we think of the men... but you know, seeing more women just in their discoveries is something that probably can be helpful.''
Based upon Esme's suggestion, highlighting more women physicists such as in lectures, seminars, newsletters, etc. can provide support for women. This can show them the achievements of women physicists who have made major contributions to physics historically and in the present, which can potentially improve their sense of belonging and persistence in physics.

\textbf{Provide reassurance during struggle}:
Ruby mentions how reassurance would be helpful especially when she does not feel confident in herself. For example, not doing well on an exam would be an opportunity to reassure students because:
``Sometimes it's scary and sometimes it makes you feel like \textit{am I in this and not able to do this?}, and a lot of the times, my confidence is definitely shaken up by a test grade or something along those lines...''
She also says that instructors should create a safe environment to get problems wrong because:
``it's okay to not always get the right answer...and that doesn't mean you're stupid and that doesn't mean that you shouldn't do what you're passionate about...''
Thus reframing mindsets about wrong answers as an opportunity to learn would be useful in addition to instructors reassuring students that they do belong in physics and that it is okay to not do well on an exam.

\textbf{Improve office hours}: 
Ruby also provides several suggestions for instructors to improve their classrooms, one of them is to improve the perception and purpose of office hours. She says,
``making a more accepting environment and comfortable environment to make mistakes and to not always know the answer...
so that people can feel confident, even when they don't know everything.''
She also calls for instructors to dismantle the idea that students should be knowledgeable about the questions they ask. She says,
``that's on the professors to make abundantly clear that it's okay, if you don't know this, it's okay, if you have a question, even if you think that this question might not be a good question, there's no such thing as a bad question.''
Thus, the responsibility is on the instructor to create this environment that feels safe to ask questions.
According to Ruby, this is a problem with the physics culture, that ``
people are terrified to even try to answer a question in the fear of getting it wrong...''
Physics culture usually rewards geniuses and creates intimidation for those who don't identify as geniuses, so dismantling this notion such that students feel supported and safe enough to go to office hours and ask questions they do not know answers to would be very helpful.
Ruby also adds that instructors should be approachable to students so that students are not as afraid to ask them questions.

\textbf{Encourage peer collaboration}:
Another suggestion Ruby provides is in regard to collaboration with peers. She wishes that there were more opportunities to work with peers because: 
``that would also set us up a lot better for once we enter the real world and we do work with other people, so I think having an environment for that would be really, really great.''
Not only would this be beneficial in introducing students to each other and forming strong connections that could support them throughout their program, but it would teach students how to collaborate with each other, as Ruby mentions.

Chloe adds to this suggestion:
``if you're not talking to people, then you're not breaking any of your preconceptions about them, so I think more engaging classroom dynamics would be good, not just for the learning, but for really like people getting to know each other on a level beyond just what they look like.''
In her opinion group work would also be beneficial so that students can get to know each other, like Ruby mentioned, but also so that students can break down their own assumptions about students (and potential biases against them).

\textbf{Restructure in-class groups}:
Chloe, who is an introvert and has difficulty breaking barriers to collaborate with peers, suggests that instructors should change how groups are formed when it comes to in-class groups, such as:
``assigning people to different groups and having rotations where people aren't always stuck with the same group but also have enough time working with other people to break down these barriers, to force people to become friends...'' Rather than keeping the same groups throughout the semester, rotating groups would help students meet each other.
``[if] teachers have control over what the format of the classroom is and they can decide if it will just be lectures, or if they will work with groups inside the classrooms.''
She justifies this by saying that instructors have control over how they structure their classes, so they can choose to incorporate group work in their classrooms and decide how groups are formed.

Paulina, who has felt isolated in physics due to the intersection of both gender and race, adds that rotating partners is better compared to being in small groups because:
``it takes away the fear of having to keep up with all these different people who might know each other and might not know you or might have different areas of strength or weaknesses. And so it's hard to kind of keep track with all of this. With one person, it's a more direct conversation, and you can slowly build comfortability around them.''
Being partnered up with another peer may reduce cognitive load and anxiety associated with meeting multiple new people.
This may also be useful for students with multiple visual marginalized identities that can impact how others perceive them, to just talk to one student at a time, rather than manage working with multiple people who can potentially team up and stereotype them or ignore them like Paulina was experiencing.

Paulina also says that being able to communicate with her instructors about whether or not she feels comfortable with her partner would be important. She explains,
``It would just be a conversation that you can have [with] the professor and then that kind of also alleviates the stress of engaging with someone who perhaps you haven't had a positive experience with before.''
It seems from Paulina negative experiences, a lot of the onus is on the student to bring this issue up. However, Paulina suggests that the instructor should make it clear that students can switch partners or provide feedback especially if they do not have good experiences with their partners. This might make it easier for students to switch partners.

\textbf{Reinforce disciplinary actions}:
Paulina emphasizes that instructors have responsibility in addressing inequities in their classroom. Specifically she refers to how students interact with each other. She also mentions that instructors have to be aware of any issues in their classrooms in order to properly intervene and take action:
``It has to start off with the professor, I guess, because since the professor is the one that's kind of in charge of engaging the students with one another, if they're not aware of it, the problem where women in STEM don't feel, women of color don't feel as if they have an accurate voice in the conversations being had and there's no way for them to personally intervene and frame their classroom to encourage the voices of women of color.'' 
Again, we see a student emphasize that instructors have to be the ones who enact change. Paulina says that to do so, they have to be aware of what is going on in their classrooms, and with their students, otherwise how could they create change? She also says that they have to be aware of the issues in their classes in order to accurately discuss them.

\textbf{Check-in with students}:
Paulina shares that it would be beneficial for instructors to check-in with students especially if some students struggle with large group settings. She describes what these check-ins would look like:
``having one on one meetings, asking students how they feel in the space and [whether] they feel comfortable, they feel their questions are being answered, not just from the professor, but what's their interactions with students?''
Paulina also adds that this can create a better relationships between the instructor and students.
These types of conversations would help instructors understand how their students experience their classrooms and what their interactions with peers are like.

\textbf{Engage in conversations with students about inequities}:
Following her suggestion for instructors to take action in their classrooms, Paulina also suggests that having workshops involving both instructors and students would be useful to address big issues. According to her, this would bring a ``proper amount of attention'' to these issues in physics.
She highlights that
``that's when change can happen, I think it has to be something that is continuous and is made an apparently big deal because if it's just kind of like something under the rug or it's again only discussed once or twice, maybe in passing, then, most people would just not really take it to heart...'' 
Here she calls for workshops to be a space where issues in physics can be highlighted as current issues for the department to work on, rather than having them be mentioned once and then them being forgotten or not addressed.

Paulina also recalls some advice from someone she met about their instructor taking time out of class to have students engage with each other. That person conveyed to her that rather than focus on solving physics problems, students
``engage with each other, and we would kind of provide a space where students can share their experiences if they wanted to...
students and professors would engage with the fact that there is a lack of equity in the classroom for students of color and so having these, I guess, in-class workshops or in-class discussions, made it...[feel] as if it was important or situation that required a class time...''
Paulina shares that having conversations frequently is useful because:
``not only [do] the students begin to expect these kinds of conversations, but as time went on, you can see that there is more openness to ask more questions all around the classroom...''
So the practice of normalizing important conversations might help other students feel comfortable asking for advice or insights.
For example, Paulina says that attending a PWI, there are many White students who might not be aware that they have biases or harmful perceptions of students of color that is
``harmful to our learning experiences and so having a chance for us to have these continuous conversations allowed for them to come into the next workshop for example being \textit{Okay, well, I learned this from the last workshop, how can I expand on this, or how can I help other conversations and ask questions if I don't really know how my actions or thoughts affect the students of color in the classroom}.''
She said that these workshops can provide an opportunity for people from dominant groups, such as White people, to be vulnerable about not understanding certain experiences and wanting feedback.

\section{Broader Discussion and Brief Comparison to Previous Studies}
We now situate this work by 
summarizing how it compares to previous qualitative studies related to experiences of undergraduate women majoring in physics. We start with brief summaries of the two studies conducted by Johnson (2020) and Santana and Singh (2023).

\subsection{Johnson's 2020 study}

In 2020, Johnson conducted open-ended interviews in a small physics department at a small liberal arts college in the US with 6 undergraduate women majoring in physics,
and a focus group consisting of third year physics students. She asked undergraduate women, during the individual interviews, 
questions like ``Tell me your life story in physics." Four physics faculty in the department were also interviewed but we will not focus on that here.
Out of the individual interviews with undergraduate women, three were women of color and three were White women. At this institution, 25\% of physics majors are women. 
For focus groups, she asked students about their trajectory in physics, what students like about their physics classes, what could be improved, and what ideas they might have for attracting more women to physics.

We chose to compare the study discussed here in a medium-size physics department at a small liberal arts college with this study because of the overwhelmingly positive physics culture that both students and faculty talked about \cite{johnson2020intersectional}. We believe this physics department can serve as a good model for an equitable and inclusive physics environment where all students thrive and feel safe \cite{johnson2020intersectional}. 

\subsection{Santana and Singh's 2023 study}

Santana and Singh conducted semi-structured, empathetic interviews with 16 undergraduate women physics and astronomy majors at a large research university in the US. At the time when the interviews were conducted, this number of women interviewed represents approximately 60\% of the women 
students in the physics and astronomy department (23\% women). Six out of 16 interviews were analyzed which included a wide range of experiences.
The goals and the protocols were similar as this current study.

Findings from this study revealed a masculine physics culture that did not support undergraduate women \cite{santana2023effects}. In particular, this study revealed that women had many negative interactions with their male peers and male instructors.
the findings point to an inequitable and non-inclusive physics learning environment unlike the one described in Johnson’s study.
One reason we chose this study for this comparative analysis is that it highlights undergraduate women's experiences 
in a physics department at a large research university in the US that may represent a more prototypical physics department.

\subsection{Comparing the three studies}
This comparison focuses on analyzing all four Domains of Power and evaluating the relative experiences of women students in the physics learning environments. We believe the study discussed here in a medium-size physics department at a small liberal arts college shows a physics learning environment which is significantly better than the one at the large research university \cite{santana2023effects} but not as equitable and inclusive as the environment in the small physics department in Johnson's study \cite{johnson2020intersectional}, which can be taken as a gold standard that all physics departments should aspire to be like. 
Thus, these two previous studies are useful for making comparisons.

First, we will discuss the interpersonal domain, which relates to individuals communicating with each other, or in our context, where students interact with peers and faculty members.
In Johnson's study \cite{johnson2020intersectional}, undergraduate women described their physics peers as ``friendly," ``super nice," and approachable. They perceived other physics students as partners in their learning so ``why wouldn't they help?" Students also got to know faculty members in their department. 
On the other hand, in Santana and Singh's 2019 study \cite{santana2023effects}, undergraduate women perceived their male peers to be condescending and to have many biases against women.
In this current study, we saw that many women felt that their male peers were condescending and some male peers perceived the women to be incapable of doing physics without their help. Some of the women tried and managed to avoid their male peers. However, some of the women noted that they did not have extremely negative interactions with their male peers.

In regard to the interpersonal domain between students and instructors, Johnson \cite{johnson2020intersectional} found that women perceived their instructors as accessible, nice, helpful. They even said that their instructors asked ``how are you feeling? how are things going?" These check-ins definitely showed them that faculty care about their students.
Contrasting these relationships with our previous study \cite{santana2023effects}, women noted that their physics instructors were generally very condescending and also had biases, which reflected stereotypes about women not being smart enough to excel in physics. The undergraduate women also reported some experiences as microaggressions from their male peers and instructors. Some women even said that male faculty set examples of bad behavior that male students then tried to emulate.
In our current study, women perceived some of their faculty as very helpful if they went to their office hours. Most of the women had overall positive opinions of their instructors. They also noted that the faculty that were either their research advisors or academic advisors were very helpful and they had positive experiences with them.

In terms of the cultural domain, Johnson's study showed an extremely positive physics culture, very atypical of many traditional physics department and classroom settings. Women students loved being around the physics building, thus showing that the physical space was welcoming and fostered an environment that emphasized working together.
On the other hand, in our previous study \cite{santana2023effects}, most women did not want to interact with male peers and the culture in spaces such as study groups or the undergraduate physics lounge, where men dominated, was very toxic. Both peers and instructors were reinforcing toxic and stereotypical physics cultures that included beliefs about physics only being for smart people, women not being able to do well in physics, etc.
In this study, some of the women chose not to work with male peers. Specifically, Paulina felt that she could not work with anyone because of how they perceived her as a woman of color and due to the lack of strong commitment to work collaboratively even when she was placed in a group for the first time at the time the interviews were conducted. She also specifically felt that cliques were formed, especially if students already knew each other (e.g., from previous classes), so it was difficult for her to work effectively in groups. From Paulina's perspective, we also saw how her identity as a woman of color impacted these interactions and she perceived her peers to negatively stereotype her, which further alienated her and made her anxious.

In Johnson's study, she described the physics structural domain to be interactive and collaborative. For instance, faculty structured their classes to incorporate group work during lectures and faculty also guided students on how to work effectively in groups \cite{johnson2020intersectional}. 
In our previous study, we noted that physics courses were taught traditionally, with few faculty incorporating active-learning strategies. In particular, faculty did not typically incorporate collaborative elements such as group work during their lectures.
In this study, many of the women did describe working in groups, but did occasionally work alone. Thus, there is some evidence that some instructors had group work embedded in their lectures. However, many of them also called for more emphasis on peer collaboration by their instructors and effective norms for peer collaborations. For example, a few students suggested working in pairs as opposed to small groups in order to not be marginalized (e.g., Paulina) as well as to reduce cognitive load and get to know their peers well.

The disciplinary domain is essential to this study, as this directly falls under faculty responsibilities. In Johnson's study, there were many instances of faculty intervening and ``reprimanding students who failed to work equitably in groups" and reacted to students reinforcing stereotypical ideologies of attributes of physicists.
Our previous study at the large research university showed that professors did not do much disciplining. In fact, some women noted that they were the ones showcasing the inequitable physics culture, demonstrating it to students, especially male students, who then continued on to perpetuate it in other contexts. The undergraduate women did make several suggestions which were focused upon disciplinary actions.
In this study, we saw few instances in which instructors addressed inequities perpetuated by male students. In particular, none of these types of issues were addressed in a meaningful way to have sustained impact beyond a short period or upheld by continuous disciplinary actions. A few women said that the instructor addressed the situation once and then students returned to similar behaviors.

The HELPIEE framework emphasizes that those in the position of power, e.g., the instructors, have the power to create equitable and inclusive learning environments for students. Instructors can use this power to create positive change and set the norms in the various related Domains of Power in Johnson's framework. Applying this to the disciplinary domain, instructors should be the ones to enforce and uphold good behavior using the suggestions for instructors provided by the women in this study. They are also the ones to establish the structure of the classroom (structural domain), i.e., creating opportunities for students to work together as well as setting norms, which could work to dismantle preconceptions or stereotypes (thus changing the cultural domain). This in turn can improve how students interact with each other, and can help students see each other as partners in learning, as opposed to using stereotypes to judge others (affecting the interpersonal domain).

\section{Conclusions}
In this study, we interviewed seven undergraduate women majoring in physics in a medium-size physics department who attended a small liberal arts PWI in the US. Through these interviews, we learned about their early experiences in physics during high school, how their family supported them in physics, e.g., by providing emotional and academic support, their interactions with their college peers and faculty, reflections about how certain intersections of their identity may have affected their experiences in physics, advice based upon what has helped them persist in physics, and suggestions for faculty that would improve their experiences.

We note that the women's family played a large role in supporting them, from an early age including during high school. They also provided a range of opportunities, which fostered interest in physics and science. However, beyond emotional support, most of these undergraduate women were unable to rely on family for academic support during college. They mainly credited some of their faculty as providing support for them. This might be the case as all of the women moved away from home to attend this college and their instructors and other physics faculty were closer in proximity to provide support especially in physics. Thus, based upon their experiences, instructors play a very important role in the women's college experiences in physics once they are away from home.
 
In particular, we see that their relationships with and perceptions of the faculty they regularly interacted with were generally overall positive. The women mentioned that women faculty were especially inspiring to them. Some of the faculty who served as academic advisors or instructors also encouraged a few of the women like Paulina or Ruby to major in physics. At least one of these faculty incorporated discussions about the experiences of marginalized people in physics. Despite these positive perceptions and experiences, some of the women noted that faculty often did not take actions or were not aware of equity and inclusion issues in their own classes. Thus, these undergraduate women had negative experiences with their peers which persisted without any intervention or with only superficial interventions by the instructor when such issues were explicitly brought out.

Again, based upon Domains of Power and HELPIEE frameworks, we emphasize that physics instructors have a lot of power in their classrooms to empower students and so it is important that they use that power to address inequities in the classroom. For example, several of the women had experiences in which their male peers talked to them in a condescending way, took credit for their work, or in Paulina's case, made her feel excluded from group work. Instructors appeared to be rarely aware of these experiences in their classroom or addressed such issues superficially when informed about them so that they persisted. Also, even if they communicated to their students that they could choose to switch partners, the women still felt like the onus was on them to bring up negative interactions with their peers to their instructor's attention.

We also saw that several women reflected on various aspects of their identities that affected their experiences negatively, which may be useful for faculty to consider. We encourage instructors to continue to contemplate upon and consider various intersections of students' identity and how those may impact students in physics learning environments since students are not a monolith and have a wide range of backgrounds and experiences. Our findings could be useful for faculty professional development workshop leaders in educating and informing instructors on how to better support diverse groups of students.

In our interviews, we also asked the women about things that would help improve the physics learning environments. Some of them provided advice based upon experiences that have personally helped them. However, we note that their advice may not be suitable for all students and some of them may put the onus on the marginalized students to change in order to improve their experiences, rather than focusing on the systemic issues that caused those negative experiences and the need for the physics culture to change to support them. Moreover, students who are marginalized in physics due to multiple marginalized identities, such as Paulina, may not benefit from simply being advised to change their mindset about their struggle unless the physics culture is improved to support them as some of the white women in this study suggested has helped them.

The women also had many suggestions for instructors, many of which were about ways to change the structure of the classrooms. Others were more focused on the responsibility of the instructor. We encourage physics faculty to consider these suggestions as many of them do not require monetary resources, but require serious reflection and time commitment on their part. In other words, these suggestions are not impossible to implement, but may require planning and collaboration amongst faculty and other experts in positions of power. Faculty also do not need to be physics education researchers to carry out these changes; instead, they must be committed towards equity and inclusion.
We also note that because instructors have power in several domains, they are responsible for setting the tone for how students are treated and establish classroom norms.
In addition, for many physics programs, men make up the majority of both physics majors and physics faculty. Hence, they have both power and influence to make change.
This change is necessary and being neutral signals both a privilege and polite oppression, which is  a phrase coined by Ferretti \cite{ferretti2018neutrality}.

With regard to the Domains of Power and HELPIEE frameworks, the power in physics learning environments directly refers to the fact that physics faculty and other individuals can make the learning environments equitable and inclusive and take action and educate those who violate the equitable and inclusive norms. Enforcing rules, such as those described by a code of conduct, and even maintaining them through the repetitive use of discussions, training and disciplinary action can potentially be useful. However, at this small college where we conducted the interviews, from the perspective of these undergraduate women, the instructors were not aware of students not treating other students equally or if these issues were pointed out, they addressed these issues superficially in class as opposed to in a way that would make sustained difference. Thus, their actions in the disciplinary domain, if any, were relatively minor. 
The cultural domain in the Domains of Power framework pertains,
e.g., to how members of the physics community think about physics, and how those beliefs are reinforced by other members. 
According to these undergraduate women, the physics culture at this college is much like a prototypical physics culture, e.g., where doing physics is associated with brilliance.

As discussed in the preceding section, if we were to rank this physics department relative to the other two mentioned here \cite{johnson2020intersectional, santana2023effects} in terms of the physics culture, it would be somewhere in the middle. However, at a small liberal arts institution with medium-size or small physics department, the culture may be easier to change because there are fewer faculty who have to be on the same page, whereas at a large research university or a fragmented department, it would be more difficult to dismantle an established physics culture with, for example, 40+ physics faculty \cite{santana2023effects}. In such a large physics department with inequitable physics culture, there would need to be a ``critical" number of faculty who are committed towards equity and inclusion in a physics department, who could work to change the physics culture \cite{santana2023effects}. With enough faculty support and resources, we believe that there can be a sustained and systemic effort in physics departments of any size and demographic distribution to create equitable and inclusive learning environments and support traditionally marginalized students, although some initiatives could require more effort than others.

In summary, through the experiences of undergraduate women presented here, we illustrated how they navigated and perceived the physics culture and engaged in doing physics. How power interacts in domains such as interpersonal, cultural, and disciplinary is especially important as they are entangled with each other.
Because physics instructors and other administrators have large roles, influence, and power, according to the HELPIEE framework, they should take action, rather than expecting students to change or adapt to the physics culture. These actions should focus on traditionally marginalized students, e.g., they can focus on implementing interventions or workshops to dismantle stereotypes in the classroom. There is also need for systemic institutional changes that would require more faculty and administrative support. In summary, 
physics faculty should take these findings and suggestions seriously and use their positions of power to implement positive changes and create a supportive environment for traditionally marginalized students.

\section{Limitations and Future Directions}
One limitation of the study is that participants, although they constituted one third of the women physics majors at the time, were self-selected, i.e., they volunteered. This may have limited other types of experiences from other women in the physics program at this institution. Future work can investigate women's experiences at other physics departments of different sizes at various types of institutions including minority serving institutions. This can also ensure that their perspectives are represented in the broader physics education research community. It would also be valuable to carry out similar investigations in other countries with undergraduate women majoring in physics.

We also note that the protocol questions were not explicitly shaped around the Domains of Power framework. However, many of the questions aligned with both Standpoint Theory and the Domains of Power frameworks. For example, we asked women questions regarding their interactions with peers and faculty, class structure, and physics culture. In answering the protocol and follow-up questions, the women reflected on the instructors' actions which can be mapped onto different Domains of Power.
We also note that Johnson used Domains of Power framework to analyze data in her 2020 study \cite{johnson2020intersectional} and asked questions similar to our questions instead of asking explicit questions shaped around the Domains of Power framework.
\section*{Acknowledgements}
We are very grateful to the faculty member who helped us recruit participants for this study and to all of the participants who volunteered for this study.

\section{Appendix}

\subsection{Protocol Questions}

\begin{table}[tb]
\caption{\label{tab:questions}Example of the relevant protocol questions used to probe undergraduate women's experiences.}
\begin{ruledtabular}
\begin{tabular}{l}
Example Protocol Questions \\
\hline
\multirow{1}{42em}{Can you tell me about your experiences with science, math, and physics in high school? How does college science compare to learning in high school?} \\
\\
\multirow{1}{42em}{What contributed to your decision to major in physics?} \\
\multirow{1}{42em}{Did you work with other people on homework, or studying for tests, in your first year in physics? What about other classes?}\\
\\
\multirow{1}{42em}{Do you typically work with people of the same gender as you, a different gender, or a mixture? Which do you prefer?}\\
\\
\multirow{1}{42em}{Can you describe an occasion where you struggled in the physics program? How do you react when you face challenges like this?} \\
\\
\multirow{1}{42em}{Do you feel that your contributions are valued by your peers and instructors?}\\
\multirow{1}{42em}{Do you feel respected as a physics person by your peers and instructor?}\\
\multirow{1}{42em}{Have you had any mentors or people who inspired you to do physics, in school, at college, or in your outside life?}\\
\\
\multirow{1}{42em}{Do you feel supported by your instructors in your physics courses?}\\
\multirow{1}{42em}{Do you think your experiences in physics have been different because of your gender? If so, how?} \\
\multirow{1}{42em}{What advice would you give to young women who are considering majoring in physics?}\\
\multirow{1}{42em}{What could the physics department do to make the environment less chilly, or to remove barriers, or to make the playing field level for men and women?}\\
\\
\end{tabular}
\end{ruledtabular}
\end{table}

\subsection{General codes}

\clearpage

\subsection{Definitions and examples for general codes.}
\begin{longtable}{lllll}
\caption{Section numbers for analytic themes (AT), subthemes, and corresponding definitions and examples for general codes. Analytic themes are located in the main document (section IV: Findings and Discussion).}\label{tab:examples}\\
AT & Subtheme & General Code & Definition & Example from Data \\
\hline
\endfirsthead

\hline
\multicolumn{5}{l}{Continued from previous page} \\
\hline
\endhead

\hline
\endfoot

\hline
\endlastfoot
A & \multirow{1}{6em}{Family's role in providing support} & \multirow{1}{6em}{Family providing academic support} & \multirow{1}{12em}{Ways in which family provided insight about science or help with homework} & \multirow{1}{15em}{``whenever I had issues with like homework and wanted help, I would always...go to her and she would help me out...''} \\
& & & & \\
& & & & \\
& & & & \\
A & \multirow{1}{6em}{Family's role in providing support} & \multirow{1}{6em}{Family providing exposure to science} & \multirow{4}{12em}{Ways in which family members, whose jobs were related to STEM, would discuss science topics, bring them to their place of work, or create a positive experience related to STEM} &  \multirow{4}{15em}{``...my grandpa was very into science and at a young age, he would always do these math puzzles with me and I just found [it] really interesting...''} \\
& & & & \\
& & & & \\
& & & & \\
& & & & \\
& & & & \\
& & & & \\
& & & & \\
A & \multirow{1}{6em}{Positive high school experiences} & \multirow{1}{6em}{Positive interactions with teachers} & \multirow{3}{12em}{Interactions students had with their teachers that were positive. For example, teachers were supportive, kind, helpful, etc.} & \multirow{3}{15em}{``I always felt they treated me like I was smart and capable and deserved to be there...''} \\
& & & & \\
& & & & \\
& & & & \\
& & & & \\
A & \multirow{1}{6em}{Positive high school experiences} & \multirow{1}{6em}{Positive interactions with peers} & \multirow{1}{12em}{Interactions students had with their peers that were positive. For example, peers were inspirational, driven, etc.} & \multirow{3}{15em}{``...it was nice being in a class where I had friends there and then they're all very driven.''}\\
& & & & \\
& & & & \\
& & & & \\
& & & & \\
& & & & \\
& & & & \\
A & \multirow{1}{6em}{Negative high school experiences} & \multirow{1}{6em}{Negative interactions with teachers} & \multirow{1}{12em}{Interactions students had with their teachers that were negative. For example, teachers showed biases towards boys' opinions, did not focus on learning, or were unprofessional.} & \multirow{1}{15em}{``...even if my hand was up or if another girl's hand [was]...[my teacher] would always call on the male students before the female students''} \\
& & & & \\
& & & & \\
& & & & \\
& & & & \\
& & & & \\
& & & & \\
& & & & \\
A & \multirow{1}{6em}{Negative high school experiences} & \multirow{1}{6em}{Negative interactions with peers} & \multirow{1}{12em}{Interactions students had with their peers that were negative. For example, peers were focused on grades as opposed to learning.} & \multirow{1}{15em}{``...I felt like everyone around me was thinking a lot about GPA and not really caring about the learning so I got a little discouraged..''}\\
& & & & \\
& & & & \\
& & & & \\
& & & & \\
& & & & \\
B & \multirow{1}{6em}{Positive interactions \& perceptions} & \multirow{1}{6em}{Positive experiences with college peers} & \multirow{1}{12em}{Interactions and opinions students had with their peers where they felt supported, safe, comfortable, working with them.} & \multirow{1}{15em}{``...people I really talk to are other girls and that has been the most comfortable for me''} \\
& & & & \\
& & & & \\
& & & & \\
& & & & \\
& & & & \\
B & \multirow{1}{6em}{Negative interactions \& perceptions} & \multirow{1}{6em}{Negative experiences with college peers} & \multirow{1}{12em}{Interactions and opinions students had with their peers where men dominated conversations, acted condescending, etc.} & \multirow{1}{15em}{``...if me and my partner had different answers, it kind of felt like he’d automatically assume that he was right, which wasn't always the case...''} \\
& & & & \\
& & & & \\
& & & & \\
& & & & \\
& & & & \\
& & & & \\
& & & & \\
& & & & \\
B & \multirow{1}{6em}{Negative interactions \& perceptions} & \multirow{1}{6em}{Negative opinions of college peers} & \multirow{1}{12em}{Negative opinions of male peers. Women believed men thought they were more capable or smarter than them.} & \multirow{1}{15em}{``...I think he thought of women in a sense, that they were less capable and more vulnerable than men, and so he wanted to be this savior...''} \\
& & & & \\
& & & & \\
& & & & \\
& & & & \\
B & \multirow{1}{6em}{Paulina as an island among her peers} & \multirow{1}{6em}{Being negatively perceived by peers} & \multirow{1}{12em}{Instances where Paulina reflects on how her intersectional identity impacts her interactions with peers.} & \multirow{1}{15em}{``there is more of a willingness to either poke fun or underestimate me because of the intersectionality of my gender and race.''} \\
& & & & \\
& & & & \\
& & & & \\
& & & & \\
B & \multirow{1}{6em}{Paulina as an island among her peers} & \multirow{1}{6em}{Being ignored by peers} & \multirow{1}{12em}{Instances where peers do not listen to or ignore Paulina. Other students choose not to engage with her.} & \multirow{1}{15em}{``...so a lot of the times I'll bring [up] points and it'll just be ignored and that's something that I've had to deal with a lot.''} \\
& & & & \\
& & & & \\
& & & & \\
& & & & \\
B & \multirow{1}{6em}{Paulina as an island among her peers} & \multirow{1}{6em}{Working alone} & \multirow{1}{12em}{Caused by being ignored by others. These include instances, e.g., where Paulina feels like working alone is the only option .} & \multirow{1}{15em}{``...no one is reaching out to each other to talk and work things out together so by default I just feel like I have to end up doing things on my own.''} \\
& & & & \\
& & & & \\
& & & & \\
& & & & \\
C & \multirow{1}{6em}{Positive interactions \& perceptions} & 
\multirow{1}{6em}{Positive interactions with instructors} & \multirow{1}{12em}{Positive interactions students had with instructors. This includes things instructors did that students thought were helpful.} & \multirow{1}{15em}{``[my instructors] have really made attempts to make their classroom environments better...''} \\
& & & & \\
& & & & \\
& & & & \\
& & & & \\
& & & & \\
& & & & \\
& & & & \\
& & & & \\
C & \multirow{1}{6em}{Positive interactions \& perceptions} & \multirow{1}{6em}{Positive opinions about instructors} & \multirow{1}{12em}{Instances when students have positive opinions about instructors. For example, they thought their instructors were caring, helpful, supportive.} & \multirow{1}{15em}{``...we're about to do a project and she's trying to make sure not to put the women in our class as the only woman in a group...and I definitely am grateful for that...''} \\
& & & & \\
& & & & \\
& & & & \\
& & & & \\
& & & & \\
C & \multirow{1}{6em}{Negative interactions \& perceptions} & 
\multirow{1}{6em}{Negative interactions with instructors} & \multirow{1}{12em}{Negative interactions students had with instructors, including being singled out.} & \multirow{1}{15em}{``...the instructor has kind of singled me out in the classroom and asked me questions continuously trying to...make sure that I'm keeping up with the class...''} \\
& & & & \\
& & & & \\
& & & & \\
& & & & \\
& & & & \\
C & \multirow{1}{6em}{Negative interactions \& perceptions} & \multirow{1}{6em}{Negative opinions about instructors} & \multirow{1}{12em}{Instances when students have negative opinions about instructors. For example, they thought instructors were not aware of what was going on in the classroom.} & \multirow{1}{15em}{``...a lot of professors don't go out of their way to look at the problems and...will assume [things are fine] or [don't] want to have to deal with anything like that.''} \\
& & & & \\
& & & & \\
& & & & \\
& & & & \\
& & & & \\
& & & & \\
D & \multirow{1}{6em}{Reflection about identities \footnote{This is also the Analytic Theme since there is only one subtheme.}} & 
\multirow{1}{6em}{Reflection on gender} & \multirow{1}{12em}{Moments where women reflected on how their gender identity impacted their experiences in physics.} & \multirow{1}{15em}{``When there's less women, I feel some of that pressure...to like be smart, if you feel like you're representing woman...''} \\
& & & & \\
& & & & \\
& & & & \\
& & & & \\
D & \multirow{1}{6em}{Reflection about identities} & 
\multirow{1}{6em}{Reflection on ethnicity/race} & \multirow{1}{12em}{Moments where women reflected on how their ethnic/racial identity impacted their experiences in physics.} & \multirow{1}{15em}{``there’s also been this expectation that I am just not up to par with them, be-
cause [of] my racial background...''}\\
& & & & \\
& & & & \\
& & & & \\
& & & & \\
D & \multirow{1}{6em}{Reflection about identities} & 
\multirow{1}{6em}{Reflection on personality} & \multirow{1}{12em}{Moments where women reflected on how their personality impacted their experiences in physics.} & \multirow{1}{15em}{``...I tend to be more introverted...outside of the classroom. I think personality does create some divides...''} \\
& & & & \\
& & & & \\
& & & & \\
D & \multirow{1}{6em}{Reflection about identities} & 
\multirow{1}{6em}{Reflection on education} & \multirow{1}{12em}{Moments where women reflected on how their prior education impacted their experiences in physics.} & \multirow{1}{15em}{``I think I didn't so much feel like isolated or alienated because of that...I think again, I've been pretty cushioned in physics.''} \\
& & & & \\
& & & & \\
& & & & \\
E & \multirow{1}{6em}{People providing emotional support} & 
\multirow{1}{6em}{Support from Family} & \multirow{1}{12em}{Ways in which family members provided emotional support or encouragement during their undergraduate degree.} & \multirow{1}{15em}{``They'd all told me, `You gotta go with your gut... because you know what you can do and you can do more than what you think you can do,' and that really stuck with me.''} \\
& & & & \\
& & & & \\
& & & & \\
& & & & \\
& & & & \\
E & \multirow{1}{6em}{People providing emotional support} & \multirow{1}{6em}{Support from Faculty Members} & \multirow{1}{12em}{Ways in which faculty members provided emotional support or encouragement during their physics trajectory.} & \multirow{1}{15em}{``...having someone that took the time to have a conversation with me and help me figure things out was really meaningful to me.''} \\
& & & & \\
& & & & \\
& & & & \\
& & & & \\
E & \multirow{1}{6em}{People providing emotional support} & \multirow{1}{6em}{Faculty as Role Models} & \multirow{1}{12em}{Ways in which women looked up to their instructors or were inspired by them.} & \multirow{1}{15em}{``He's one of the few other Latinx professors that I've seen on campus... it's become very meaningful to see someone from a similar background as me kind of succeed.''} \\
& & & & \\
& & & & \\
& & & & \\
& & & & \\
& & & & \\
& & & & \\
& & & & \\
& & & & \\
& & & & \\
F & \multirow{1}{6em}{Personal Advice} & 
\multirow{1}{6em}{Involvement in the Department} & \multirow{1}{12em}{Ways in which women became involved in the physics department through leadership roles. These roles are ways that have provided a better experience for the women.} & \multirow{1}{15em}{``...I liked that once I was involved, I could then kind of make the events that I thought would be helpful so I'm glad I did that.''} \\
& & & & \\
& & & & \\
& & & & \\
& & & & \\
& & & & \\
& & & & \\
F & \multirow{1}{6em}{Personal Advice} & \multirow{1}{6em}{Persisting with a Growth Mindset} & \multirow{1}{12em}{Ways in which women adopted a growth mindset and how it has helped them persist in the physics major.} & \multirow{1}{15em}{``...what has helped me the most is learning that it's okay to not understand everything... it's about just practicing and trying and also just learning as much as you can.''} \\
& & & & \\
& & & & \\
& & & & \\
& & & & \\
& & & & \\
F & \multirow{1}{6em}{Advice for Instructors} & 
\multirow{1}{6em}{Avoid forming opinions based on students' backgrounds} & \multirow{1}{12em}{Suggestion provided by women for faculty members not to make assumptions about students based on their identity or background.} & \multirow{1}{15em}{``...a small school like [my college], where you might have professors for multiple years, you might come back to that professor that you've had your first semester and I think that's important.. for professors to know.''} \\
& & & & \\
& & & & \\
& & & & \\
& & & & \\
& & & & \\
& & & & \\
F & \multirow{1}{6em}{Advice for Instructors} & \multirow{1}{6em}{Address and dismantle biases against women} & \multirow{1}{12em}{Suggestion provided by women for faculty members to communicate and address stereotypes against women in science} & \multirow{1}{15em}{``...making it a very, very clear fact that...men are no better at physics than women.''} \\
& & & & \\
& & & & \\
& & & & \\
& & & & \\
& & & & \\
& & & & \\
& & & & \\
& & & & \\
F & \multirow{1}{6em}{Advice for Instructors} & \multirow{1}{6em}{Address lack of representation of women in physics.} & \multirow{1}{12em}{Suggestion provided by women for faculty members to address the historical lack of representation of women in physics.} & \multirow{1}{15em}{``I think it'd be great to address the fact that there are not many historical women in physics, but that does not mean that there cannot be in the future...''} \\
& & & & \\
& & & & \\
& & & & \\
& & & & \\
F & \multirow{1}{6em}{Advice for Instructors} & \multirow{1}{6em}{Provide examples of contribution of women in physics} & \multirow{1}{12em}{Suggestion provided by women for faculty members to provide examples of contributions of women and 
need for more support for women in physics.} & \multirow{1}{15em}{``...seeing more women just in their discoveries is something that probably can be helpful.''} \\
& & & & \\
& & & & \\
& & & & \\
& & & & \\
& & & & \\
F & \multirow{1}{6em}{Advice for Instructors} & \multirow{1}{6em}{Provide reassurance during struggle} & \multirow{1}{12em}{Suggestion provided by women for faculty members to recognize and reassure students regarding their performance and struggle.} & \multirow{1}{15em}{``...making it more clear that it's okay to not always get the right answer...and that doesn't mean you're stupid...''} \\
& & & & \\
& & & & \\
& & & & \\
& & & & \\
F & \multirow{1}{6em}{Advice for Instructors} & \multirow{1}{6em}{Improve office hours} & \multirow{1}{12em}{Suggestion provided by women for faculty members to improve the structure and culture of office hours.} & \multirow{1}{15em}{``...get rid of this notion that students are supposed to come in already knowing everything...that they're asking about.''} \\
& & & & \\
& & & & \\
& & & & \\
F & \multirow{1}{6em}{Advice for Instructors} & \multirow{1}{6em}{Encourage peer collaboration} & \multirow{1}{12em}{Suggestion provided by women for faculty members to encourage students to work together and form community.} & \multirow{1}{15em}{``...I'd love to be able to talk with other students in a classroom as well as outside the classroom and learn from them and I don't think there's enough of that.''} \\
& & & & \\
& & & & \\
& & & & \\
& & & & \\
F & \multirow{1}{6em}{Advice for Instructors} & \multirow{1}{6em}{Restructure in-class groups} & \multirow{1}{12em}{Suggestion provided by women for faculty members to change how groups are formed.} & \multirow{1}{15em}{``...having rotations where people aren't always stuck with the same group but also have enough time working with other people to break down these barriers..''} \\
& & & & \\
& & & & \\
& & & & \\
F & \multirow{1}{6em}{Advice for Instructors} & \multirow{1}{6em}{Reinforce disciplinary actions} & \multirow{1}{12em}{Suggestion for instructors to discipline students when they are not working equitably.} & \multirow{1}{15em}{``if they're not aware of [issues in the classroom], there's no way for them to personally intervene....''} \\
& & & & \\
& & & & \\
& & & & \\
F & \multirow{1}{6em}{Advice for Instructors} & \multirow{1}{6em}{Check-in with Students} & \multirow{1}{12em}{Suggestion for instructors to check on the well-being of students.} & \multirow{1}{15em}{``asking students how they feel in the space and [whether] they feel comfortable, they feel their questions are being answered...what [are] their interactions with students...''} \\
& & & & \\
& & & & \\
& & & & \\
& & & & \\
& & & & \\
F & \multirow{1}{6em}{Advice for Instructors} & \multirow{1}{6em}{Engage in conversations with students about inequities} & \multirow{1}{12em}{Suggestion for instructors to have conversations with students about inequities in physics.} & \multirow{1}{15em}{``...having conversations where...students and professors would engage with the fact that, there is a lack of equity in the classroom for students of color...''} \\
& & & & \\
& & & & \\
& & & & \\
& & & & \\
\end{longtable}

\clearpage
\bibliography{ref.bib}
\end{document}